\documentclass[rmp,aps,nofootinbib,endfloats,amsmath,amssymb]{revtex4}
\usepackage{graphics}
\usepackage[latin1]{inputenc}
\usepackage[active]{srcltx}
\usepackage{setspace}
\usepackage{graphicx}
\usepackage{dcolumn}
\usepackage{bm}
\usepackage{epsfig}

\begin{document}
\title{On Newton's Third Law and its Symmetry-Breaking Effects}
\author{Mario J. Pinheiro}
\email{mpinheiro@ist.utl.pt}
\affiliation{Department of Physics, Instituto Superior Tecnico,
Av. Rovisco Pais, 1049-001 Lisbon, Portugal}
\affiliation{Institute of Plasmas and Nuclear Fusion, Av. Rovisco Pais, 1049-001, Lisbon, Portugal}

\begin{abstract}
The law of action-reaction, considered by Ernst Mach the cornerstone of physics, is thoroughly used to derive the conservation laws of
linear and angular momentum. However, the conflict between momentum conservation law and Newton's third law, on experimental and theoretical grounds, call for more attention.
We give a background survey of several questions raised by the action-reaction law and, in particular, the role of the physical vacuum is shown to provide an appropriate framework to clarify the occurrence of possible violations of the action-reaction law. Then, in the framework of statistical mechanics, using a maximizing entropy procedure, we obtain an expression for the general linear momentum of a body-particle. The new approach presented here shows that Newton's third law is not verified in systems out of equilibrium due to an additional entropic gradient term present in the particle's momentum.
\end{abstract}

\maketitle
\tableofcontents

\section{INTRODUCTION}
\label{sec:intro}

The law of action-reaction, or Newton's third law~\cite{Newton}, is
thoroughly used to derive the conservation laws of
linear and angular momentum. Ernst Mach considered the third law as
``his most important achievement with respect to the
principles"~\cite{Mach_01,Jammer}. However, the reasoning used
primarily by Newton applies to point particles without structure and
is not concerned with the motion of material bodies composed with a
large number of particles, in or out of thermal equilibrium.

Ernst Mach sustained that the concept of mass and Newton's third law
were redundant; that in fact it should be enough to define
operationally the mass of a given body as the unit of mass to be
sure that ``If two masses 1 and 2 act on each other, our very
definition of mass asserts that they impart to each other contrary
accelerations which are to each other respectively as
2$:$1"~\cite{Mach_01}. Yet philosophy has delivered us extraordinary
new insights to a basic understanding of the underlying physics of
force. For example, F\'{e}lix Ravaisson~\cite{Ravaisson} in the XIX
century sustained that within the realm of the inorganic world
action-equals-reaction; they are the same act perceived by two
different viewpoints. But in the organic world, whenever more
complex systems are at working, ``Ce n'est pas assez d'un moyen
terme indiff\'{e}rent comme le centre des forces oppos\'{e}es du
levier; de plus en plus, il faut un centre qui, par sa propre vertu,
mesure et dispense la force"~\footnote{``It is not enough an indifferent middle
agent, like the center of opposed forces acting on the lever; it is
necessary an agent that, by its own virtues, measure and control the
force" (translated by the author).}. So, there is in Nature the need of an ``agent" that control
and deliver the action from one body to another and this is, as we
will see, the role of the physical vacuum, or just barely the
environment of a body.

We can find in Cornille~\cite{Cornille} a review of applications of
the action-reaction law in several branches of physics. In addition,
Cornille~\cite{Cornille_2003} introduced the concepts of spontaneous force (obeying to
Newton's third law) and stimulated force (which violates it), clarifying the nature os spontaneous emission with interest to electron accelerators and lasers.

In this paper we review major aspects of action-to-reaction law in the frame of classical mechanics and electrodynamics, as described by the skew rank 2 field tensor $F_{\mu \nu}=\partial_{\nu} A_{\mu}-\partial_{\mu}A_{\nu}$ which is not affected by the gauge transformation $A_{\mu} \mapsto A_{\mu}+ie\partial_{\mu} \Lambda$, invariant under the symmetry $U(1)$ group (group of all rotations about a given axis) with Abelian commutation relations (extension to non-abelian $SU(2)$ group~\cite{Edmonds,Khvorostenko,Barrett} and higher symmetry forms~\cite{Baum} may lead to {\it symmetry breaking} and the existence of longitudinal electric fields, and these subjects are out of the scope here). Also, we intend to show that, in general, for any system out of
equilibrium with velocity-dependent entropy terms, Newton's third law
is violated. The need for re-examination of this problems is
pressing since long-term exploitation of the cosmos face serious
problems due to outdated spacecraft technologies mankind
possess. And this principle is fundamental and instrumental in understanding physics.

Sec. II offer methodological notes related to the action-reaction law, as it appears in mechanics and electrodynamics. Sec. III discusses the possible role of physical vacuum as a third agent which might explain action-to-reactions law violations. Secs. IV and V discusses the intrinsic violation of Newton's third law for systems out-of-equilibrium. Sec. VI presents the conclusions that follow logically from the previous discussion.

\section{BACKGROUND SURVEY}

The usual derivation of the laws governing the linear and angular
momenta presented in textbooks is as follows. The equation of
motion for the $i$th particle is given by:
\begin{equation}\label{eq0}
\mathbf{F}_i + \sum_{j \neq i} \mathbf{F}_{ij} = \frac{d
\mathbf{p}_i}{d t},
\end{equation}
which is Newton's second law, and where $\mathbf{F}_i$ denotes the {\it external} force acting on the $i$ particle (due to an external source), $\mathbf{F}_{ij}$ represents the {\it internal} force exerted on the
particle $i$ by the particle $j$, and $\mathbf{p}_i=m_i \mathbf{v}_i$. For a single particle, if the force $\mathbf{F}$ derives from a potential function $U(\mathbf{r},t)$, then the equation of motion is written as
\begin{equation}\label{eq0a}
m\frac{d \mathbf{v}}{dt}=-\nabla U.
\end{equation}
Multiplying by the velocity $\mathbf{v}$, we have:
\begin{equation}\label{eq0b}
m\frac{d\mathbf{v}}{dt} \cdot \mathbf{v} = -\nabla U \cdot \mathbf{v}.
\end{equation}
From Eq.~\ref{eq0} we may conclude that, if we assume the validity of the action-to-reaction-law, Eq.~\ref{eq0b} can be written in the form of the law of conservation of energy:
\begin{equation}\label{eq0c}
\frac{d}{dt} \left( \frac{1}{2}m\mathbf{v}^2 + U \right)=0.
\end{equation}
Thus, we can infer that the validity of the law of conservation of energy depends on two assumptions: i) the external force is conservative, $\mathbf{F} = -\nabla U$; ii) action-to-reaction law is observed, e.g.: for two particles, $\mathbf{F}_{12}=-\mathbf{F}_{21}$. We can talk of {\it mutual interaction} only when Newton's third law is verified. In the same line of thought, we define {\it closed system} as one that does obey to Newton's third law; an {\it open system} is one that is acted by {\it external} force(s) that {\it by definition} does not obey to Newton's third law. When external forces are zero, we say that the system is {\it closed}, or {\it isolated}. These statements will be instrumental in clarifying different situations (see also Ref.~\cite{Cornille}).

In the case of central forces the
relation $\mathbf{F}_{ij}=-\mathbf{F}_{ji}$ is indeed verified, in fact a
manifestation of Newton's third law. Summing up all the
particles belonging to the system, we have from Eq.~\ref{eq0}:
\begin{equation}\label{eq01}
\sum_i \mathbf{F}_i=\sum_i \frac{d \mathbf{p}_i}{d t}.
\end{equation}
Podolsky~\cite{Podolsky} called our
attention to the discrepancies obtained when directly using Newton's
second law, or by using instead the invariance of the lagrangian under
rotations. In the case of non-central forces, like a system
subject to a potential function of the form $V=r^{-1} \cos
\vartheta$, we might expect a deviation from Newton's third law.
Indeed, angle-dependent potentials, long-range (van der Waals)
forces, describe rigorously the physical properties of molecular
gases. One can but wonder from which mechanism it comes the unbalance of forces.

We might expect that thermodynamics and statistical mechanics both provide a more complete description of macroscopic matter. The internal energy and, in
particular, the average total energy of a system $\overline{E}=\sum_i
U_i$, which includes summing up all the particles constituting the
system and all storage modes, plays a fundamental role together with
an equally fundamental, although less understood entity, the
entropy of the system. Interesting enough, a microscopic model of friction shown that
the irreversible entropy production is drawn from the increase of
Shannon information~\cite{Diosi_02}.

This question is related to the fundamental one, still not answered
by physicists and biophysicists: how chaos in various natural systems can
spontaneously transform to order? The observation of various
physical and biological systems shows that a feedback is onset
according to: ``The medium controls the object-the object shapes the
medium"~\cite{Tsyganov_91}. At the microscopic level, it have been study a large class
of systems generating directed motion through the interaction of a
moving object with an inhomogeneous substrate periodically
structured~\cite{Popov_02}. This is the ratchet-and-pawl principle.

It is well-known the apparent violation of the Newton's third law at microscopic scale which occurs, e.g., when two equal charged bodies having equal
velocities in magnitude and opposing directions cross each other. The
Lorentz's force actuating on both electric charges do not cancel each other
since the magnetic forces do not actuate along a common line (see also
the Onoochin's paradox~\cite{McDonald}). The paradox is solved
introducing the electromagnetic momentum $[\mathbf{E} \times
\mathbf{H}]/ c^2$ (values in SI units will be used throughout the text)~\cite{Keller_42}.

In the domain of astrophysics the same problem appears again. For instance,
based on unexplained astrophysical observations, such as the high
rotation of matter around the center of the galaxy, it was proposed a
modification of Newton's equations of dynamics~\cite{Milgrom_83},
while more recently a new effect was reported, about the possibility
of a violation of the Newton's second law with bodies experimenting
spontaneous acceleration~\cite{Ignatiev_07}. In the frame of statistical mechanics, studying the effective forces exerted between two fixed big colloidal particles immersed in a bath
of small particles, it has been shown that the nonequilibrium force field is
nonconservative and violates the action-to-reaction law~\cite{Likos_03}.

An ongoing debate on the validity of electrodynamic force law is
still raging~\cite{Wesley 1996}, with experimental evidence that Biot-Savart law does
not obeys action-to-reaction law (see Ref.~\cite{Graneau_82,Graneau_01,Gerjuoy:49}
and references therein). The essence of the problem stands on two
different laws that exist in magnetostatics, giving the force between
two infinitely thin line-current elements $d \mathbf{s}_1$ and
$d\mathbf{s}_2$ through which pass currents $i_1$ and $i_2$. The
{\it Amp\`{e}re's law} states that this force is given by:
\begin{equation}\label{eq2}
d^2 \mathbf{F}_{2,A}=-\frac{\mu_0 i_1 i_2}{4
\pi}\frac{\mathbf{r}_{12}}{r^3_{12}}[ 2(d\mathbf{s}_1 \cdot
d\mathbf{s}_2) - \frac{3}{r^2_{12}}(d\mathbf{s}_1 \cdot
\mathbf{r_{12}})(d\mathbf{s}_2 \cdot \mathbf{r}_{12}) ].
\end{equation}
This means that the force between two current elements depends not
only on their distance, as in the inverse square law, but also on
their angular position (in particular, implicating the existence of
a longitudinal force, experimentally confirmed by Saumont~\cite{Saumont_68} and Graneau~\cite{Graneau_87}, and discussed by Costa de Beauregard~\cite{Beauregard_93} and Ref.~\cite{Pinheiro 2009}). The other force, generally considered, is given
by the {\it Biot-Savart law}, also known as the Grassmann's equation
in its integral form:
\begin{equation}\label{eq3}
d^2 \mathbf{F}_{2,BS} = - \frac{\mu_0 i_1 i_2}{4 \pi}
\frac{1}{r^3_{12}} [ (d\mathbf{s}_2 \times (d\mathbf{s}_1) \times
\mathbf{r}_{12})].
\end{equation}
Here, $\mathbf{r}_{12}$ is the position vector of element 2 relative
to 1. While Amp\`{e}re's law obeys Newton's third law, Biot-Savart
law does not obey it (e.g.,
Ref.~\cite{Christodoulides_88,Graneau_94,Valverde1,Valverde2}). The theory developed by Lorentz was criticized by H. Poincar\'{e}~\cite{Poincare_00}, because it sacrificed action-to-reaction law.

The problem of linear momentum of stationary system of charges and
currents is faraway from the consensus too. Costa de
Beauregard~\cite{Costa_67} pointed out a violation of the
action-to-reaction law in the interaction between a current loop $I$
flowing on the boundary of area $\mathbf{A}$ with moment
$\mathbf{\mathfrak{M}}=I \mathbf{A}$ and an electric charge,
concluding that when the moment of the loop changes in the presence
of an electric field, a force must act on the current loop, given by
$\mathbf{F}=[\mathbf{E} \times \mathbf{\mathfrak{\dot{M}}}]/c^2$.
Shockley and James~\cite{Shockley_67} have attributed $\mathbf{F}$ to a
change in the ``hidden momentum" $\mathbf{G}_l=-[\mathbf{E} \times
\mathbf{\mathfrak{M}}]/c^2$, carried within the current loop by the
steady state power flow, necessary to balance the divergence of the
Poynting's vector. The total momentum is $\mathbf{p}=\mathbf{G}_l +
\mathbf{G}_b$, where $\mathbf{G}_b=m<\mathbf{\dot{r}_{CM}}>$ is the body momentum associated with the center of mass $m$
~\cite{Shockley_68,Haus_68}. In particular, it was
shown~\cite{Shockley_68} that the ``hidden linear momentum" has as
quantum mechanical analogue the term $\mathbf{\alpha} \cdot
\mathbf{E}$, where $\mathbf{\alpha}$ are Dirac matrices appearing in
the hamiltonian form $\widehat{H} \psi=i \hbar
\partial \psi/\partial t$, where $\widehat{H}=-ic \hbar \mathbf{\alpha} \cdot
\nabla \cdot$ is the hamiltonian operator (e.g., Ref.~\cite{Sakurai}). Although certainly an important issue, the concept of ``hidden momentum" needs  further clarification~\cite{Boyer_05}.

Calkin~\cite{Calkin_71} has shown that the net linear momentum for
any closed stationary system of charges and currents is zero, and it can
be written:
\begin{equation}\label{eq4}
\mathbf{P}=\int d^3 r \mathbf{r} \left(\frac{\dot{u}}{c^2} \right)=M
\mathbf{r}_{CM},
\end{equation}
where $u$ is the energy density, $M$ is the total mass $M=\int d^3
r (u/c^2)$, and $\mathbf{r}_{CM}$ is the radius vector of the center
of mass. He has shown, however, that the linear mechanical momentum
$\mathbf{P}_{ME}$ in a static electromagnetic field is nonzero and
is given by:
\begin{equation}\label{eq5}
\mathbf{P}_{ME}=-\int d^3 r \rho \mathbf{A}^T.
\end{equation}
Here, $\mathbf{A}^T$ denotes the transverse vector potential given
by $\mathbf{A}^T=(\mu_0/4\pi) \int d^3 r \mathbf{J}/r$. Eq.~\ref{eq5} shows that $\rho \overrightarrow{A}$ is a measure of momentum per unit volume.

Similar conclusions were obtained by Aharonov {\it et al.}~\cite{Aharonov_88}
showing, in particular, that the neutron's electric dipole moment in a
external static electric field $\mathbf{E}_0$ experiences a force
given by $m\mathbf{a}=-(\mathbf{v} \cdot \nabla)(\mathbf{v} \times
\mathbf{E}_0)$. The experimental verification of the Aharonov-Casher effect would confirm total momentum conservation when occurs
interactions of magnets and electric charges~\cite{Goldhaber_89}.

Breitenberger~\cite{Breitenberger} discusses thoroughly this
question, showing the delicate intricacies behind the subject,
pointing out the conservation of canonical momentum and the
``extremely small" effect of magnetic interactions, making an analysis based on the Darwin's lagrangian, derived in 1920~\cite{Darwin_20}.
Boyer~\cite{Boyer_06} applying the Darwin's lagrangian to the system
of a point charge and a magnet, has shown that the center-of-energy
has uniform motion. The Darwin's lagrangian is correct to the order
$1/c^2$ (remaining Lorentz-invariant) and the procedure to obtain it
eliminates the radiation modes and, thus, describes the interaction
of charged particles in the frame on an action-at-a-distance
electrodynamics. However, it can lead to unphysical
solutions~\cite{Bessonov_99}.

Hnizdo~\cite{Hnizdo_92} has shown that at nonrelativistic
velocities, the Newton's third law is verified in the interactions between
current-carrying bodies and charged particles because the
electromagnetic field momentum is equal and opposite to the hidden
momenta, hold by the current-carrying bodies; the mechanical
momentum of the entire {\it closed} system is conserved. Hnizdo also has shown that, however, the field angular momentum in a system is not
compensated by hidden momentum, and thus the mechanical angular
momentum is not conserved alone, but had to be summed with the
field angular momentum, in order to become a conserved quantity.

In fact, the ``magnetic current force", produced by magnetic charges
that ``flow" when magnetism changes, given by
$\mathbf{f}_m=\varepsilon_0 \mathbf{E} \times (\mathbf{\dot{B}} -
\mu_0 \mathbf{\dot{H}})$\cite{ShockleyJames} is the ``Abraham term",
appearing in the Abraham density force $\mathbf{f}_A$ which differs
from the Minkowsky density force $\mathbf{f}_M$ through the
equation:
\begin{equation}\label{eq6}
\mathbf{f}_A = \frac{\partial }{\partial t} [\mathbf{g}^M - \mathbf{g}^A].
\end{equation}
Here, $\mathbf{g}^M=[\mathbf{D} \times \mathbf{B}]$ is the Minkowsky momentum density of the field and $\mathbf{g}^A=[\mathbf{E} \times \mathbf{H}]/c^2$ is the Abraham momentum density.

\section{INTERACTION WITH THE VACUUM}

Although Newton's third law of motion apparently does not complies for
some situations, action and reaction are likely to occur by
pairs and a kind of accounting balance such as
$\mathbf{F}=-\mathbf{F'}$ holds.

According to the Maxwell's theorem, the resultant of $\mathbf{K}$
forces applied to bodies situated within a closed surface $S$ is
given by the integral over the surface $S$ of the Maxwell stresses tensor:
\begin{equation}\label{Eq7}
\int_S \mathbf{T}(n) dS = \int_V \mathbf{f} d v = \mathbf{K}.
\end{equation}
Here, $\mathbf{f}$ is the ponderomotive forces density and $d v$
is the volume element. The vector $\mathbf{T}(n)$ under the integral
in the left-hand side (lhs) of the equation is the tension force
acting on a surface element $dS$, with a normal $\mathbf{n}$
directed toward the exterior and it is assumed the integration is done over a constant volume. In cartesian coordinates, each
component of $\mathbf{T}(n)$ is defined by
\begin{equation}\label{eq8}
T_x(n)=t_{xx} \cos (n,x) + t_{xy} \cos (n,y) + t_{xz} \cos (n,z),
\end{equation}
with similar expressions for $T_y$ and $T_z$. The 4-dimensional
electromagnetic momentum-energy tensor (in flat spacetime) of rank 2 (with respect to the three-dimensional rotations) is a generalization of the 3-dimensional
(Maxwell's) stress tensor $\sigma_{\alpha \beta}$ (in cgs-Gaussian units):
\begin{equation}\label{eq8a}
\sigma_{\alpha \beta}=\frac{1}{4\pi} \left[ E_{\alpha} E_{\beta} + B_{\alpha} B_{\beta} -\frac{\delta_{\alpha \beta}}{2} (\mathbf{E}^2+\mathbf{B}^2) \right].
\end{equation}
The indices $\alpha$ and $\beta$ refer to the coordinates $x$, $y$, and $z$, and $\delta_{\alpha \beta}$ is the Kronecker delta. Since Maxwell, the stress is one of the field properties, in addition to energy, power and momentum, consistent with experimental observations and widely used in numerical field solutions. Usually fields and matter interact, and the stress-energy tensor must be a summation of their respective contributions, $T=T^{matter}+T^{fields}$. For convenience, we may here recall that for a viscous fluid, the stress-energy tensor is given by~\cite{Landau2}:
\begin{widetext}
\begin{equation}\label{eq8b}
T_{ij}^{fluid}=p\delta_{ij}+\rho v_i v_j -\eta\left( \frac{\partial v_i}{\partial x_j}+\frac{\partial v_j}{\partial x_i}-\frac{2}{3}\delta_{ij}\frac{\partial v_l}{\partial x_l} \right) + \zeta \delta_{ij} \frac{\partial v_l}{\partial x_l}.
\end{equation}
\end{widetext}
Here, $\eta$ and $\zeta$ are the viscous coefficients. For an isotropic body, the stress tensor $\sigma_{ij}^{sb}$ is given by~\cite{Landau TE}:
\begin{equation}\label{eq8c}
\sigma_{\alpha \beta}^{sb}=K u_{\gamma \gamma } \delta_{\alpha \beta} + 2 \mu \left( u_{\alpha \beta}-\frac{1}{3}\delta_{\alpha \beta} u_{\gamma \gamma} \right),
\end{equation}
where $u_{ij}$ is the deformation tensor; $K$ and $\mu$ are, resp., the moduli of compression and rigidity.

If electric charges are inside a conducting
body in vacuum, in presence of electric $E$ and magnetic $H$ fields,
then Eq.~\ref{Eq7} must be modified to the form:
\begin{equation}\label{eq9}
\int_S \mathbf{T}(n) dS - \mathbf{K} = \int_V \frac{1}{4 \pi c} \left(
\frac{\partial [\mathbf{E} \times \mathbf{H}]}{\partial t} \right) d
v.
\end{equation}
In the right-hand side (r.h.s.) of the above equation it now appears the
temporal derivative of $\mathbf{G}=\int \mathbf{g}d \Omega$, the
electromagnetic momentum of the field in the entire volume contained
by the surface $S$ (with $\mathbf{g}$ denoting its momentum density). The integrals have to be done over a sufficiently large volume $V(\mathbf{r},t)$ bounded by a closed surface $S(\mathbf{r},t)$ containing all particles and fields.

In the case the surface $S$ is filled with a homogeneous medium
without true electric charges, Abraham proposed to write the
following equation:
\begin{equation}\label{eq10}
\int_S \mathbf{T}(n) dS = \frac{\partial }{\partial t} \int_V
\left( \frac{\varepsilon \mu}{4 \pi c} [\mathbf{E} \times \mathbf{H}] \right) d
v,
\end{equation}
with $\varepsilon$ and $\mu$ the dielectric constant of the medium
and its magnetic permeability, and assuming constant volume of integration.

As remarked by Selak {\it et al.}~\cite{Selak 1989} and Cornille~\cite{Cornille_2003}, if the volume of integration is not constant
Eq.~\ref{eq9} should be written under the form
\begin{equation}\label{eq10a}
\mathbf{K} + \frac{\partial}{\partial t} \int_{V(t)} \frac{1}{4 \pi c} \left(
[\mathbf{E} \times \mathbf{H}] \right) d v = \int_{S(t)} \mathbf{T}_{eff}(n) dS,
\end{equation}
where the effective stress-energy tensor is given by
\begin{equation}\label{eq10b}
\mathbf{T}_{eff} = \mathbf{T}-c \mathbf{P}_r.
\end{equation}
Here, $\mathbf{P}_r = \frac{1}{4 \pi c}(\mathbf{E} \times \mathbf{B} + P\mathbf{E})$, and $P$ denoting the polarization vector (see Ref.~\cite{Cornille_2003}). This transformation is necessary because it is not permissible to substitute a convective time derivative for an Eulerian time derivative when we have a non constant and finite volume of integration. The wrong assessment of this problem may lead to contradictions when, e.g., a moving vacuum-plasma boundary is modeled~\cite{Bellan 1986}. This problem was discussed in Ref.~\cite{Pinheiro 2007}, where it has been shown that with the convective derivative, the Lorentz's equation is just an outcome of Maxwell's equations, and not a necessary condition to complete the system of fundamental equations of the electromagnetic field.

Eq.~\ref{eq10} can be written on the form of a general conservation
law:
\begin{equation}\label{eq11}
\frac{\partial \sigma_{\alpha \beta}}{\partial x_{\beta}} -
\frac{\partial g_{\alpha}}{\partial t} = f_{\alpha}
\end{equation}
where $\alpha, \beta=1,2,3$, $\sigma_{\alpha \beta}$ is the stress tensor,
$g_{\alpha}$ is the momentum density of the field, and $f_{\alpha}$
is the total force density. After some algebra, this equation can
take the final form (e.g., Ref.~\cite{Ginzburg76}):
\begin{equation}\label{eq12}
\frac{\partial \sigma_{\alpha \beta}}{\partial x_{\beta}}=
f_{\alpha}^L + \frac{1}{4 \pi c} \frac{\partial}{\partial t}
[\mathbf{D} \times  \mathbf{B}]_{\alpha} + f'_{m,\alpha}.
\end{equation}
Here, $f'_m$ is the total force acting in the medium (see
Ref.~\cite{Ginzburg76}), $\mathbf{f}^L=\rho_e
\mathbf{E}+\frac{1}{c}[\mathbf{j} \times \mathbf{B}]$ is the Lorentz
force density with $\rho_e$ denoting the charge density and
$\mathbf{j}$ the current density. The second
term in the r.h.s. of the above equation, could possible be called {\it
vacuum-interactance} term ~\cite{Clevelance} - in fact, it is the Minkowski
term. According to an interpretation of Einstein and Laub
~\cite{Laub}, when integrating the above equation over all space, the
derivative over the stress tensor gives a null integral, and the Lorentz's
forces summed over all the universe must be balanced by the quantity
$\int_{\infty} \varepsilon_0 \mu_0 \frac{\partial [\mathbf{E} \times
\mathbf{H}]}{\partial t} dV$ in order to be verified Newton's third law~\cite{Cornille_2003}. It is important to remark that the field momentum $[\mathbf{D} \times \mathbf{B}]$ is equivalent to $\rho \mathbf{A}$, the first term is related to the stress-tensor representation, while the second one is related to the ``fluid-flow" representation~\cite{Carpenter_89}. Hence, the last remark, drives us to the Machian view of the origin of mass which had fascinated Einstein to such a degree that he sought to build his general theory of relativity on that ground.
Einstein gave the first published reference to Mach's principle in Ref.~\cite{Einstein 1912}: ``...the entire inertia of a point mass is the effect of the presence of all other masses, deriving from a kind of interaction with the latter". In this sense, Mach's principle (supported by Einstein during the early years of his work on general relativity, but not in his later period) seeks to restore action-to-reaction law in the entire universe.

Of course, field, matter and physical vacuum together form a closed
system and it is usual to catch the momentum conservation law in the
general geometric form~\cite{Thirring,Landau2,Lee}:
\begin{equation}\label{eq13}
\frac{\partial (T_{\alpha \beta}^{Field} + T_{\alpha \beta}^{Matter}
+ T_{\alpha \beta}^{Vacuum})}{\partial x_{\beta}} = 0.
\end{equation}
The table~\ref{table1} shows the different expressions for the
energy-momentum tensors of Minkowksy, $T_{\alpha,\beta}^M$ and
Abraham, $T_{\alpha,\beta}^A$.

\begin{table*}
\caption{\label{table1} Expressions for the energy-momentum tensors
of Minkowksy $T_{ik}^M$ and Abraham $T_{ik}^A$,
using $i,k=1,2,3,4$; $\alpha, \beta=1,2,3$; $x_1=x$, $x_2=y$, $x_3=z$, $x_4=ict$. The
Poynting's vector is $\mathbf{S}=[\mathbf{E} \times \mathbf{H}]$ and
the energy for a system at rest is $w=\frac{1}{8 \pi} (\epsilon E^2
+ \mu H^2)$.}
\begin{ruledtabular}
\begin{tabular}{ccc}
Minkowsky        &  Abraham \\ \hline
  $T_{ik}^M= \left(\begin{array}{cc}
                       \sigma_{\alpha,\beta} & -ic\mathbf{g}^M \\
                        -\frac{i}{c}\mathbf{S} & w
                      \end{array} \right)$
  &  $T_{ik}^A=\left( \begin{array}{cc}
                       \sigma_{\alpha,\beta} & -ic\mathbf{g}^A \\
                        -\frac{i}{c}\mathbf{S} & w
                      \end{array} \right) $\\
  $ \mathbf{g}^M = \frac{\epsilon \mu}{c^2}[\mathbf{E} \times \mathbf{H}]$
  & $\mathbf{g}^A = \frac{1}{c^2}[\mathbf{E} \times \mathbf{H}]$ \\
\end{tabular}
\end{ruledtabular}
\end{table*}

The general relation between Minkowski and Abraham momentum, free of
any particular assumption, holding particularly for a moving medium,
is given by:
\begin{equation}\label{eq14}
\mathbf{P}^M = \mathbf{P}^A + \int \mathbf{f}^A dt d v.
\end{equation}

For clearness, we shall distinguish the following different parts of
a system: i) the body carrying currents and the currents themselves
(the structure, for short, denoted here by $\mathfrak{K}$), ii)
fields, and iii) the physical vacuum (or the medium).

On the theoretical ground exposed above, the impulse transmitted to the
material structure should be given by the following equation:
\begin{equation}\label{eq15}
\mathbf{P}^{\mathfrak{K}} = \int \mathbf{f}^A dt d v = \mathbf{P}^M -
\mathbf{P}^A.
\end{equation}
Here, $\mathbf{f}^A$ denotes the Abraham's force density~\cite{Abraham1,Abraham2,Pfeifer 2007}:
\begin{equation}\label{eq16}
\mathbf{f}^A = \frac{\varepsilon_r \mu_r -1}{4 \pi c} \frac{\partial
[\mathbf{E} \times \mathbf{H}]}{\partial t}.
\end{equation}

\begin{figure}
  \includegraphics[width=3.5 in]{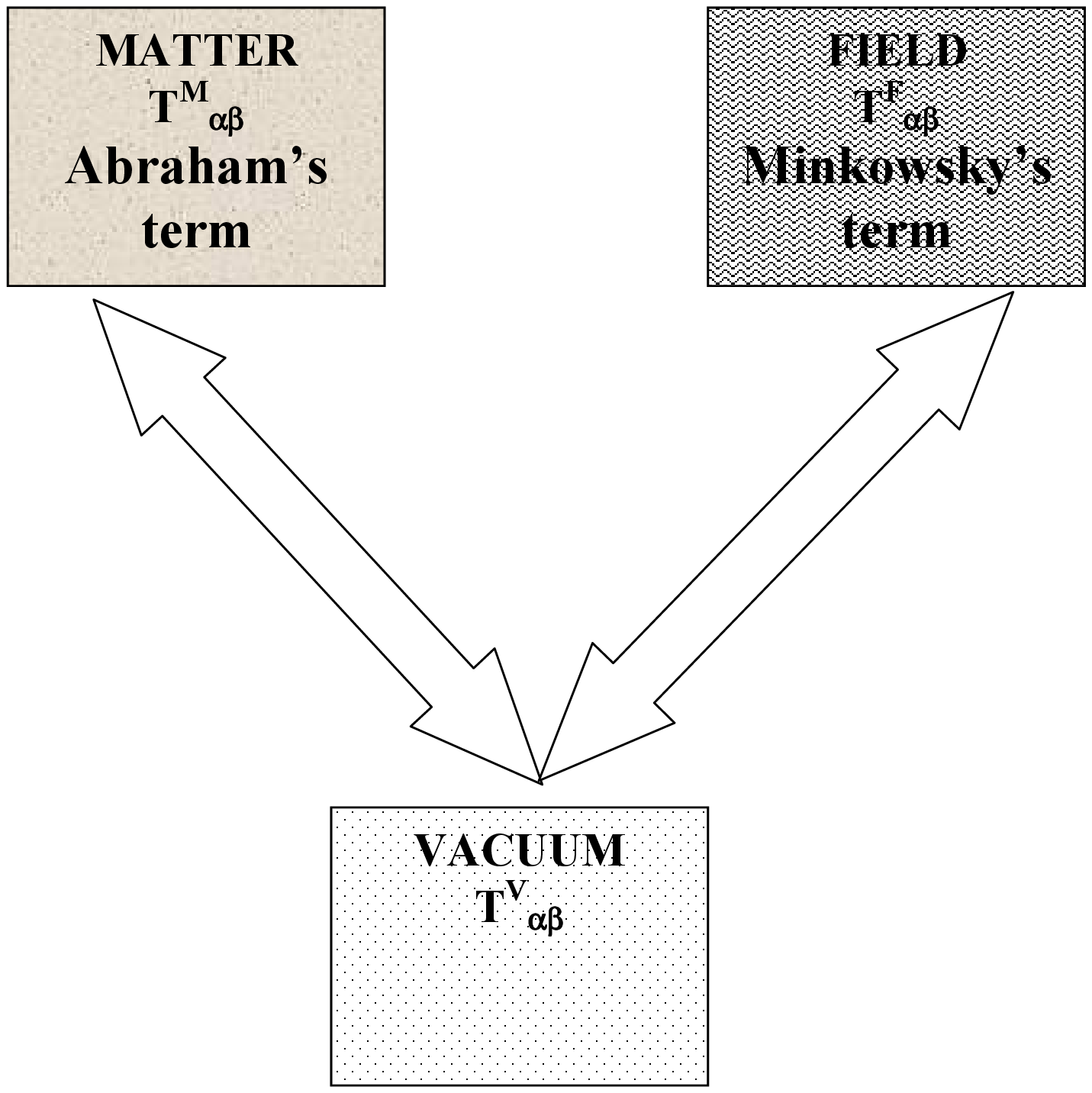}\\
  \caption{Conservation law for the closed system: Matter + Field + Physical Vacuum.}\label{fig1}
\end{figure}

This is in agreement with experimental data ~\cite{Jones} and was
proposed by others ~\cite{Gordon,Tangherlini}. As this force acts over the medium, it is expected nonlinearities related to the
behavior of the dielectric to different applied frequencies,
temperature, pressure, and large amplitudes of the electric field,
when a pure dielectric response of the material is no longer
proportional to the electric field (e.g., see Ref.~\cite{Waser} on this
topic).


As is well known, Maxwell's classical theory introduces the idea of
a real vacuum medium. After being considered useless by Einstein in his
special theory of relativity, the ``ether" (actually replaced by the
term {\it vacuum} or {\it physical vacuum}) was rehabilitated by
Einstein in 1920 ~\cite{Einstein22}. In fact, general theory of
relativity describes space as possessing physical properties by means of ten
functions $g_{\mu \nu}$ (see also ~\cite{Ginzburg87}). According to
Einstein, \begin{quote} The ``ether" of general relativity is a
medium that by itself is devoid of {\it all} mechanical and
kinematic properties but at the same time determines mechanical (and
electromagnetic) processes.
\end{quote}

Dirac felt the need to introduce the idea of ``ether" in quantum
mechanics ~\cite{Dirac}. In fact, according to quantum field theory,
the particles can condense in vacuum giving rise to space-time dependent
macroscopic objects, for example, of ferromagnetic type. Besides,
stochastic electrodynamics has shown that the vacuum contains
measurable energy, called zero-point energy (ZPE), described as
turbulent sea of randomly fluctuating electromagnetic fields. Quite
interestingly, it was recently shown that the interaction of atoms
with the zero-point field (ZPF) guarantees the stability of matter
and, in particular, the energy radiated by an accelerated electron
in circular motion is balanced by the energy absorbed from the ZPF
~\cite{Kozlowski}. An attempt to replace a field by a finite number
of degrees of freedom was accomplished by Pearle~\cite{Pearle_71}. In this
theory, a set of $N$ particles are supposed do not interact directly with
each others, but interact directly with a number of dynamical
variables (called the ``medium") carrying the ``information" from one
particle to another.

Graham and Lahoz have made three important
experiments~\cite{Lahoz1,Lahoz2,Lahoz3}. While the first experiment
provided an experimental observation of Abraham force in a
dielectric, the second one has provided evidence of a reaction
force which appears in magnetite. The third one, gave the first
evidence of free electromagnetic angular momentum created by
quasistatic and independent electromagnetic fields $E$ and $B$ in
physical vacuum ~\footnote{According to Graham and Lahoz, cited in
~\cite{Lahoz3}, ``According to Maxwell-Poynting ideas, the last
(Minkowski's) term in [our Eq.1] can be interpreted as a local
reaction force acting on charges and currents when the vacuum
surrounding them is loaded with electromagnetic momentum."}. Whereas the referred paper by Lahoz {\it et al.}
provided experimental evidence for Abraham force at low frequency
fields, it still remains to gather evidence of its validity at
higher frequency domain, although some methods have been presently
outlined~\cite{Antoci}.

In view of the above, we will write the ponderomotive force density
acting on the composite body of arbitrarily large mass (formed by
the current configuration and its supporting structure) in the form (here in SI units):
\begin{equation}\label{eq18}
\rho \frac{\partial \mathbf{V}}{\partial t} = \nabla \cdot \overleftrightarrow{T} -
\frac{\partial}{\partial t} \left( \varepsilon_0 \mu_0 [\mathbf{E}
\times \mathbf{H}] \right).
\end{equation}
Here, $\overleftrightarrow{T}$ is a dyadic representation of the electromagnetic (stress) force per unit area acting on the surface $S$; $-T_{ij}$ is the momentum in the $i$ direction crossing a surface oriented in the $j$ direction, per unit area, per unit time. Eq.~\ref{eq18} and as well Eq.~\ref{eq12}, both assume that the energy and momentum density are continuously distributed over the region of space occupied by fields. This gives rise to difficulties with the problem of absorption of light, in particular, when localized discrete particles are considered. For this reason, the above described continuity equations must be written in integral form. Accordingly, integrating Eq.~\ref{eq18} over the entire volume of the structure and fields, it gives
\begin{equation}\label{eq18a}
\frac{d \mathbf{P}_{mec}}{dt} = \oint_{S(t)} \mathbf{\overleftrightarrow{T}} \cdot d \mathbf{S} -
\frac{d}{d t} \int_{V(t)}  \left( \varepsilon_0 \mu_0 [\mathbf{E}
\times \mathbf{H}] \right) d v.
\end{equation}
The last integral represents the momenta stored in the electromagnetic field. The surface integral tends towards zero when the radius $R$ tends to infinity but, when the near-field is taken into account, this may not be true, as they decrease as $R^{-2}$ (see, e.g. Ref.~\cite{Baba_00} for an analytical example), the integral tending to a finite value~\cite{Cornille_2003} since the surface elements $dS=R^2 d\Omega$ increases as $R^2$. Hence, the surface integral is not necessarily null, as stated in several textbooks~\cite{Cohen,Landau FT,Ginzburg}, but it is correctly assessed in others~\cite{Becker,Plonsey} (see also Ref.~\cite{Cornille_2003} and references therein). The stress-energy tensor constitute a powerful technique when studying problems such as levitation~\cite{Brandt 1989}, or the action of the radiation pressure exerted by light on cells, particles and atoms~\cite{Ashkin 1986}, manipulating the concentrated electromagnetic energy in sub-wavelength regions near tips, objects or surfaces.

\subsection{Examples}

\subsubsection{Force exerted on an interface between two different media}

For example, the force exerted on an interface between two different media can be obtained by integrating the stress tensor over a cylindrical surface with its base parallel to the interface and tending subsequently the height of the cylinder to zero. This force is given by:
\begin{widetext}
\begin{equation}\label{eq18aa}
f_i = \int\int_S \left[\epsilon_2 E_{2i}E_{2j} - \epsilon_1 E_{1i}E_{1j} -\frac{1}{2}\delta_{ij} \left( E_2^2 (\epsilon_2-\eta\frac{d \epsilon}{d \eta})_2 -E_1^2 (\epsilon - \eta \frac{d \epsilon}{d \eta})_1) \right) \right] dS_j,
\end{equation}
\end{widetext}
where $\epsilon$, $\mu$ and $\eta$ are, resp., the permittivity, permeability, and mass density of the medium. When considering non-uniform periodic fields of the form $\mathbf{E}(\mathbf{r},t)=\mathbf{E}_0(\mathbf{r})e^{j \omega t}$ (most experiments are conducted at optical frequencies), and using the identity $\Re(\mathbf{A})\Re(\mathbf{B})=1/2\Re(\mathbf{A}\mathbf{B}^{*})$, with $\Re$ denoting the real part, Eq.~\ref{eq18aa} may be written under the form
\begin{widetext}
\begin{equation}\label{eq18aaa}
\overline{f}_i=\frac{1}{2} \Re{\int \int \left[\epsilon_2 E_iE_j^{*}-\epsilon_1 E_{1i}E_{1j}-\frac{1}{2}\delta_{ij}(\epsilon_2 \mid E_2 \mid^2 - \epsilon_1 \mid E_1 \mid^2) \right] dS_j},
\end{equation}
\end{widetext}
where $\overline{f}$ denotes the time average as given by $\overline{f}=\lim_{T \to \infty} \int_{-T}^{T} (f)dt$.
Its application to the problem of an oscillating charge $q=q_0 e^{j \omega t}$ facing a semi-infinite dielectric, gives the following average force transmitted by the fields across the dielectric interface~\cite{Giner 1995,Richards}:
\begin{equation}\label{eq18aa1}
\overline{f}=\frac{q_0^2}{32 \pi \epsilon_0 d^2} \Re \left(\frac{\epsilon-\epsilon_0}{\epsilon + \epsilon_0} \right),
\end{equation}
where $d$ is the distance between the oscillating charge and its image.

The role of the stress-energy tensor is made comprehensible considering that the $E$ and $B$ near-fields, both take seat on the physical space and, when a charge is accelerated it occurs a bending of the lines of force, that becomes subsequently an independent physical entity, detached from the electric charge but not accelerated with the charge~\cite{Harpaz 2004,Pinheiro 2008}. The effect of the self-field on an extended charged particle it was shown do contribute to inertia~\cite{Pinheiro 2008}.

Hence, the composite body is acted on by Minkowski force in such a
way that
\begin{equation}\label{eq19}
M\mathbf{V} = \mathbf{P}^M - \mathbf{P}^A.
\end{equation}
The Minkowski momentum is transferred only to the field in the
structure and not to the structure and the field in the
medium~\cite{Skobeltsyn,Ginzburg76,Lahoz3}. In summary, to move a
spacecraft forward, the spacecraft must push ``something" backwards;
and this ``something" might be the physical vacuum. This effect was shown to be made feasible, the Abraham's force representing the reaction of the physical vacuum fluctuations to the motion of dielectric fluids in crossed electric and magnetic fluids communicating to matter velocities of the order of 50 nm$/$s~\cite{Feigel_04}, although this result was contested by van Tiggelen {\it et al.}~\cite{Tiggelen_04}. However, the resulting tiny forces produced by the electromagnetic field momentum (or the associated Poynting's vector) made it difficult to experimentally measure Abraham's force and weakens the possibility of its application in field propulsion concepts.

\subsubsection{Graham and Lahoz experiment}

Another cornerstone of electrodynamics is the equation of conservation of angular momentum (e.g., Ref.~\cite{Tsai}):
\begin{equation}\label{eq19a}
\frac{d \mathbf{L}_m}{dt}=-\frac{d}{dt} \int_{V(t)} \frac{1}{c^2} [\mathbf{r} \times \mathbf{S}] d\Omega - \oint_{S(t)} [\mathbf{r} \times \overleftrightarrow{T}] \cdot d\mathbf{S},
\end{equation}
where we assumed that the shape of $S(t)$ depends on time. Here, $\mathbf{L}_m$ is the angular momentum of the charges (matter), $[\mathbf{r} \times \mathbf{S}]/c^2$ is the field angular momentum density, and the last term on the r.h.s. is the angular momentum flux of the field with density (tensor) $[\mathbf{r} \times \overleftrightarrow{T}]$. The component $\beta$ of the surface integral can also be represented in the form $\oint \varepsilon_{\beta \gamma \delta} x^{\gamma} T^{\delta \zeta} n_{\zeta} dS$, with $\varepsilon_{\beta \gamma \delta}$ denoting the totally antisymmetric Levi-Civita symbol (normalized by $\varepsilon_{123}=1$) and $n_{\zeta}$ is the $\zeta$ component of the unit vector outward normal to the 2-dimensional surface $S$. It is worth to note that this is a governing equation similar to Eq.~\ref{eq11}. The so called Feynman's paradox~\cite{Feynman} has been experimentally reproduced by Graham and Lahoz~\cite{Lahoz3}. In their experiment the torque on a cylindrical capacitor apparently gave evidence of a reaction acting on physical (empty) space. We may notice that when the integral on stress-energy tensor is non-null, due particularly to the action of local forces, it naturally occurs violation of action-to-reaction law. This situation happens for instance with a {\it celt stone} when spun in the appropriate  direction: due to contact forces with (local) surface and the agency of terms of the kind shown in Eq.~\ref{eq8c} it results chiral (asymmetric) behavior~\cite{Bondi 1986,Moffatt 2008}. This local contact force also explains why action-to-reaction law is not obeyed when you succeed to move any system (e.g. a closed box) by appropriate motion {\it inside} the box, {\it but} with the device in contact with a surface~\cite{Provatidis 2010}, or when self-forces are induced at a mesoscopic level on single asymmetric objects~\cite{Soto_08,Buenzli 2009}. They are all {\it open systems}.

The exploration of these ideas to propel a spacecraft as an alternative to chemical propulsion has been advanced in the literature, e.g., see Refs.~\cite{Taylor,Trammel,Brito,Forward2004,GlenMurad}, and for the particular configuration of two electric dipoles the {\it first} term on the r.h.s. of Eq.~\ref{eq18a} due to the near-field may result in propulsion, see Ref.~\cite{Baba_00} for a concrete analytical example. Also, propulsion based on Maxwell's stress tensor have been proposed by Slepian~\cite{Slepian} and Corum {\it et al.}~\cite{Corum 1999}.

\section{DEDUCING THE LINEAR MOMENTUM OF A BODY ON THE BASIS OF STATISTICAL PHYSICS}

When two bodies of matter collide, the repulsive force exerted on them is
equal whenever no dissipative process is at stake. When a ball
rebound on the floor it has the same total mechanical energy before
and after the collision, except for a loss term which is due to the
fact that the bodies have internal structure. At a microscopical
level, bodies are aggregates of molecules. When the body collides,
molecules gain an internal (random) kinetic energy. Macroscopically
this generates heat, and therefore raises the system entropy. In
global terms, some fraction of heat does not return to the
particle's collection constituting the ball and the entropy of the
universe ultimately increases.

Let us consider an isolated material body composed by a great number
of macroscopic particles (let's say $N$) possessing an internal
structure with a great number of degrees of freedom (to validate the
entropy concept) with momentum $\mathbf{p}_i$, energy $E_i$ and with
intrinsic angular momentum $\mathbf{J}_i$, all constituted of
classical charged particles with charge $q_i$ and inertial mass
$m_i$. Using the procedure outlined in
Refs.~\cite{Pinheiro:02,Pinheiro:04} we can show that the entropy
gradient in momentum space is given by:
\begin{equation}\label{eq20}
\mathbf{p}_i = m_i \mathbf{v}_e + q_i \mathbf{A} + m_i
[\mathbf{\omega} \times \mathbf{r}_i] -m_i T_i \frac{\partial
\overline{S}}{\partial \mathbf{p}_i}.
\end{equation}
It was assumed that all particles have the same drift velocity and
they turn all at the same angular velocity $\omega$. The center of
mass of the body moves with the same macroscopic velocity and the
body turns at the same angular velocity~\cite{Landau2}. The last term
of Eq.~\ref{eq20} represents the gradient of the entropy in a
nonequilibrium situation and $\overline{S}$ is the transformed
function defined by:
\begin{widetext}
\begin{equation}\label{eq21}
\overline{S} = \sum_{i=1}^{N} \left\{ S_i \left[ E_i -
\frac{p_i^2}{2m_i} - \frac{J_i^2}{2I_i} -q_i V_i + q_i (\mathbf{A}_i
\cdot \mathbf{v_i})] + (\mathbf{a \cdot \mathbf{p}_i}) + \mathbf{b}
\cdot ([\mathbf{r}_i \times \mathbf{p}_i] + \mathbf{J}_i) \right]
\right\},
\end{equation}
\end{widetext}
where $\mathbf{a}$ and $\mathbf{b}$ are Lagrange multipliers.

Whenever the system is in thermodynamic equilibrium the canonical
momentum is obtained for each composing particle:
\begin{equation}\label{eq22}
\mathbf{p}_i = \mathbf{p}_{rel} + m_i [\mathbf{\omega} \times
\mathbf{r}_i] + q_i \mathbf{A}_i.
\end{equation}
Otherwise, when the system is subjected to forced constraints in
such a way that entropic gradients in momentum space do exist, then
a new expression for the particle momentum must be taken into
account, that is, Eq.~\ref{eq20}.

Summing up over all the constituents particles of a given
thermodynamical system pertaining to the same aggregate (e.g., body
or Brownian particle), we obtain:
\begin{equation}\label{eq24}
\mathbf{P} = M \mathbf{v}_e + \sum_i m_i [\mathbf{\omega} \times
\mathbf{r}_i] + Q \mathbf{A} - \sum_i m_i T_i \frac{\partial
\overline{S}}{\partial \mathbf{p}_i}.
\end{equation}
To simplify, we can assume that all particles inside the system
share the same random kinetic energy, $T_i=\zeta$:
\begin{equation}\label{eq25}
\mathbf{P} = M \mathbf{v}_e + \sum_i m_i [\mathbf{\omega} \times
\mathbf{r}_i] + Q \mathbf{A} - \zeta \sum_i \frac{\partial
\overline{S}_{ne}}{\partial \dot{r}_i},
\end{equation}
where by $\overline{S}_{ne}$ we denote the entropy when the system
is in a state out of equilibrium. The first term on the right-hand-side is the
bodily momentum associated with the motion of the center of
mass $M$; the second term represents the rotational momentum; the
third is the momentum of the joint electromagnetic field of the
moving charges~\cite{Fowles_80,Scanio_75}; finally, the last term is
a new momentum term, physically understood as a kind of
``entropic momentum" since it is ultimately associated to the
information exchanged with the medium on the the physical system
viewpoint (e.g., momentum that eventually is radiated by the charged
particle). Lorentz's equations don't change when time is reversed,
but when retarded potentials are applied the time delay of
electromagnetic signals on different parts of the system do not
allow perfect compensation of internal forces, introducing
irreversibility into the system~\cite{Ritz_08}. This is always true
whenever there is time-dependent electric or/and magnetic
fields~\cite{Jefimenko_1}. Cornish~\cite{Cornish_86} obtained a
solution of the equation of motion of a simple dumbbell system held
at fixed distance and have shown that the effect of radiation
reaction on an accelerating system induces a self-accelerated
transverse motion. Obara and Baba~\cite{Baba_00} have discussed the electromagnetic propulsion mechanism obtained from an electric dipole system, showing that the propulsion effect results from the delay action of the static and inductive near-field created by one electric dipole on the other.
These are examples of irreversible (out of
equilibrium) phenomena that do not comply with action-reaction law.

\subsection{Example}

\subsubsection{Missing Symmetry}

At this stage, we can argue that the momentum is always a conserved
quantity provided that we add the appropriate term, in order Newton's third
law can be verified. This apparent ``missing symmetry" might result because
matter alone does not form a closed system, and we need to include
the physical vacuum in order to restore lost symmetry. So, when we have
two systems $1$ and $2$ interacting via some kind of force field
$\mathbf{F}$, the reaction from the vacuum must be included as a
sort of bookkeeping device:
\begin{equation}\label{eq26}
\mathbf{F}_{12}^{matter}=-\mathbf{F}_{21}^{matter} +
\mathbf{F}^{vacuum}.
\end{equation}
We may assume the existence of a physical vacuum probably well
described by a spin-0 field $\phi(x)$ whose vacuum expectation value
is not zero:
\begin{equation}\label{eq27}
\text{vacuum} \sim \phi(x),
\end{equation}
and at its lowest-energy state to have zero 4-momentum, $k_{\mu}=0$
(e.g., Ref.~\cite{Lee}).

This new state out of equilibrium can be constrained by applying an
external force on the system (e.g., set all system into rotation
about its central axis at the same angular velocity
$\mathbf{\omega}$).

\vspace{0.5 cm}

It was shown that the entropy must increase with a small
displacement from a previous referred state ~\cite{Lavenda,Landau2}.
Considering that the entropy is proportional to the logarithm of the
statistical weight $\Omega \propto exp(\overline{S}/k_B)$ and
considering that $\overline{S}=\overline{S}_{eq}+\overline{S}_{ne}$,
we can expect an increase of the nonequilibrium entropy
$\overline{S}_{ne}$ with a small increase of the i$th$ particle's
velocity $\mathbf{v}_i=\mathbf{\dot{r}_i}$, since with an increase
of particle's speed (although in random motion) the entropy must increases altogether.
Therefore, we must always have:
\begin{equation}\label{eq28}
T \frac{\partial \overline{S}_{ne}}{\partial \mathbf{\dot{r}_i}} \geq
0, \forall i=1,...N.
\end{equation}
In conditions of mechanical equilibrium the equality must hold, otherwise condition ~\ref{eq28} can be considered a {\it
universal criterium of evolution}. Considering that the entropy is
an invariant~\cite{Rengui} there is no extra similar term when the
momentum is transferred to another inertial frame of reference.

Quite withstanding, there is an important theorem derived by Baierlin~\cite{Baierlein}
showing that the Gibbs entropy for a system of free particles with
kinetic energy $K$, density $\rho$ and absolute temperature $T$,
$S(K,\rho,T)$, is greater than the entropy associated to the same
system subject to arbitrary velocity-independent interactions $V$,
$S(K+V,\rho,T)$, such as $S(K+V,\rho,T) \leq S(K,\rho,T)$.

At the electromagnetic level, Maxwell conceived a dynamical model of
a vacuum with hidden matter in motion. As it is well-known,
Einstein's theory of relativity eradicated the notion of ``ether"
but later revived its interest in order to give some physical mean
to $g_{ij}$. Minkowski obtained as a mathematical consequence of the
Maxwell's mechanical medium that the Lorentz's force should be
exactly balanced by the divergence of the Maxwell's tensor in vacuum
$T_{vac}$ minus the rate of change of the Poynting's vector:
\begin{equation}\label{eq29}
\rho \mathbf{E} + \mu_0 [\mathbf{J} \times \mathbf{H}] = \nabla
\cdot \mathrm{T}_{vac} -\frac{\partial }{\partial t} \varepsilon_0
\mu_0 [\mathbf{E} \times \mathbf{H}].
\end{equation}
Einstein and Laub have remarked~\cite{Laub} that when Eq.~\ref{eq8} is
integrated all over the entire Universe the term $\nabla \cdot
\mathrm{T}_{vac}$ must vanish which means that the sum of all Lorentz
forces in the Universe must be equal to the quantity $\int_{\infty}
\varepsilon_0 \mu_0 \partial / \partial t [\mathbf{E} \times
\mathbf{H}] dv$ in order to comply with Newton's third law (see
Ref.~\cite{Lahoz}). But, this long range force depends on the
constant of gravitation $G$. Einstein accepted the Faraday's viewpoint on the reality of fields, and this gravitational field
according to him would propagate all over the entire space without
loss, locally obeying to the action-reaction law. But nothing can
reassure us that the propagating wave through the vacuum will be
lost at infinite distances~\cite{Brillouin_70}.
Poincar\'{e}~\cite{Poincare_01} also argues about the possible
dissipation of the action on matter due to the absorption of the
propagating wave in the context of Lorentz's theory.

The Newton's laws are valid, generally, for large scales. When the scale tends to mesoscopic level or even smaller scales, all three Newton's laws will become invalids. The Newton's third law is acceptable in most observable scales, but when scale tends to the microscopic realm or extremely large scale, difficulties with Newtonian mechanics will arise~\cite{Vujicic_1}. In particular, according to Ref.~\cite{Vujicic_1}, the third law becomes invalid for electron interaction (e.g., Onoochin's paradox). To better handle with a possible fractal nature of spacetime, El-Naschie's E-infinity theory~\cite{El Naschie_1} regards discontinuities of space and time in a transfinite way, through the introduction of a Cantorian spacetime.

By Noether's theorem, energy conservation is related to
translational invariance in time ($t \to t +a$) and momentum
conservation is related to translational invariance in space ($r_i
\to r_i + b_i$). This important theorem thus implies that the law of
conservation of momentum (not equivalent to the
action-equals-reaction principle) is always valid, while the law of
action and reaction does not always holds, as shown in the previous
examples.
Some kind of relationship must therefore exists between entropy and
Newton's third law, since it was through the
first and second law of thermodynamics combined that our main result were
obtained. This idea was verified recently through a standard
Smoluchowski's approach, and on the Brownian dynamic computer simulation of
two fixed big colloidal particles in a bath of small Brownian
particles, drifting with uniform velocity along a given direction. It
was shown that, in striking contrast to the equilibrium case, the
nonequilibrium effective force violates Newton's third law, implying
the presence of nonconservative forces with a strong
anisotropy~\cite{Likos_03}, in concordance with our Eq.~\ref{eq26}.

\section{IS IT VERIFIED THE ACTION-EQUALS-REACTION IN A OUT-OF-EQUILIBRIUM THERMODYNAMICAL SYSTEM ?}

\begin{figure}
  \includegraphics[width=3.0 in, height=4.0 in]{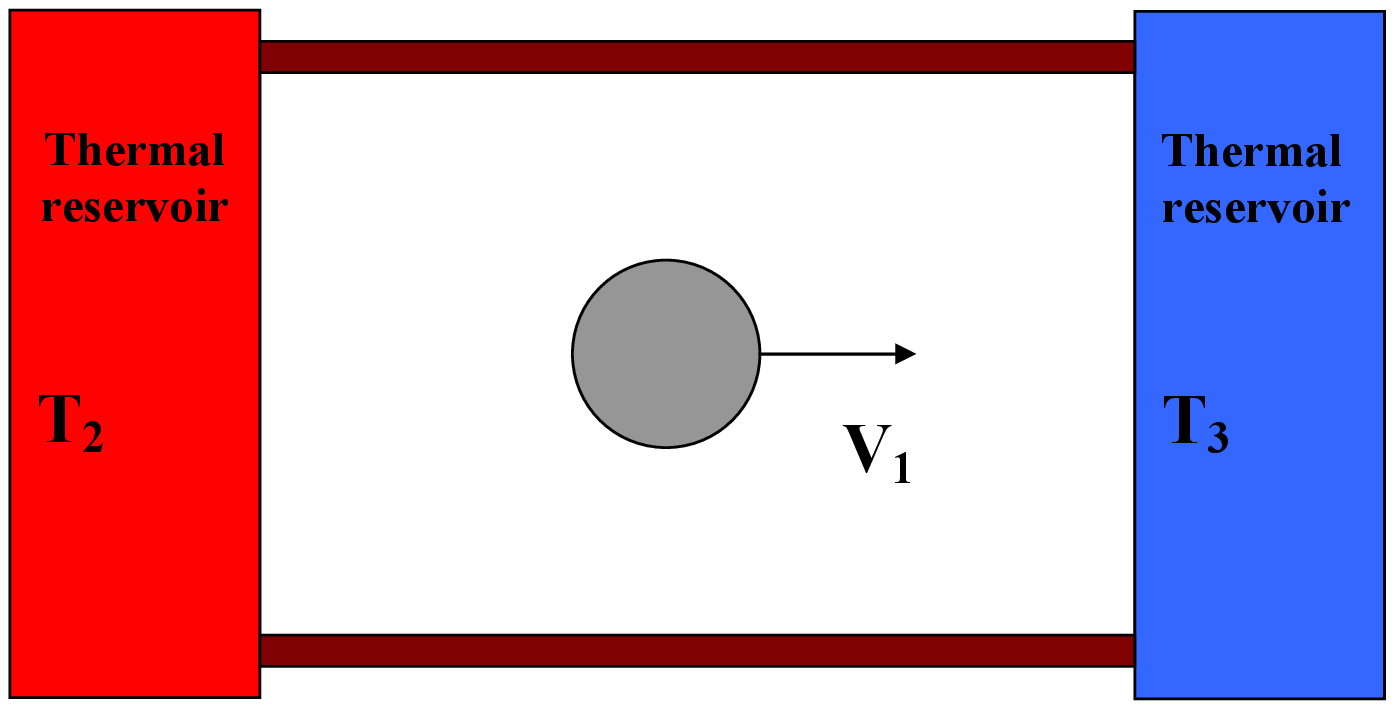}\\
  \caption{Schematic of the self-accelerated device.}\label{fig2}
\end{figure}
The maximizing entropy procedure proposed in Ref.~\cite{Pinheiro:02,Pinheiro:04} suggests the following ``gedankenexperiment", which bears some resemblance with Leo Szilard's thermodynamical engine, made up of a one-molecule fluid (e.g., Ref.~\cite{Leff}), although we are not concerned here with neguentropy issues.

\subsection{Example}

\subsubsection{Self-accelerated engine}

Let us consider a physical system consisting of a spherical body built of
$N$ number of particles closed in a box, moving along one direction (see Fig.~\ref{fig2}). The
left side is at temperature $T_2$, the right side is at
temperature $T_3$, while all the particles inside the body itself is at temperature $T_1$
(and in equilibrium with their photonic environment).
Furthermore, let us assume that both surfaces and the body particle
are all thermal reservoirs, and hence their respective temperatures do not change.
Let us suppose that the onset of nonequilibrium dynamics can be forced by some means in the previously described device.
When the particle collides with the left surface, its momentum varies according to:
\begin{equation}\label{eq30}
\delta p_{\upharpoonleft} = - mv_1'' + m v_1 + (T_3-T_1)
\partial_v S.
\end{equation}
Here, $\partial_v S$ denotes the (nonequilibrium) entropy gradient in velocity-space.
After the collision, the particle goes back to hit the right side surface
at temperature $T_3$. The momentum variation after the second
collision is given by:
\begin{equation}\label{eq31}
\delta p_{\downharpoonleft} = mv_1' - mv_1'' + (T_2 - T_1)
\partial_v S.
\end{equation}
We assume that the body attains thermal equilibrium with the
environment (which must remain at constant temperature $T_1$) fast enough before the next hit against the wall of the thermal reservoir. The total balance after a back and forth complete cycle is given by:
\begin{equation}\label{eq32}
\delta p_{\downharpoonleft} = - \delta p_{\upharpoonleft}
-\partial_v S(T_2 + T_3 - 2 T_1)=-\delta p_{\upharpoonleft}-\Delta \zeta \nabla_v S.
\end{equation}
To make it more clear, we might write Eq.~\ref{eq32} under the form:
\begin{equation}\label{eq33}
\delta p_{\downharpoonleft}=-\delta p_{\upharpoonright}-\delta p_{\upharpoonright}^{is},
\end{equation}
where we denote by $\delta p_{\upharpoonright}^{is}\equiv \Delta \zeta \nabla_v S$, the change in momentum by the physical vacuum (or, more appropriately, we should call ``inertial space").
Therefore, it is clear from the above analysis that
{\it in systems out of equilibrium} Newton's third law is not
verified, but the conservation of canonical momentum is well verified, however, as it must be according to Noether's theorem. Otherwise,
when the temperatures are equal to all thermal bath in contact,
such as $T_1=T_2=T_3$, Newton's third law is complied:
\begin{equation}\label{eq33}
\delta p_{\downharpoonleft} = - \delta p_{\upharpoonleft}.
\end{equation}
In the frame of nonlinear dynamics and statistical approach, Denisov has shown~\cite{Denisov_02} that a rigid shell and a nucleus with internal dynamic asymmetric can perform self unidirectional propulsion. Also, it seems now certain, that depletion forces exerted between two big colloidal particles in a bath of small particle, exhibit nonconservative strongly anisotropic forces that violate action-to-reaction law~\cite{Likos_03} (see also Ref.~\cite{Wang_02}). In addition, internal Casimir's forces exerted between a circle and a plate in nonequilibrium situation violates Newton's law~\cite{Soto_08}.

\subsubsection{Stimulated Emission versus Newton's Third Law}

Considering the radiation as a reservoir, Einstein~\cite{Einstein 1917} introduced master equations seeking to describe the effect of absorption, stimulated emission, and spontaneous emission processes between two levels of an atom immersed in the black-body radiation field. These equations read:
\begin{equation}\label{eq33a}
\left\{ \begin{array}{cc}
\frac{dN_b}{dt} = -A_{ba}N_b + u(\omega)(B_{ab}N_a - B_{ba}N_b) \\
\frac{dN_a}{dt} =  A_{ba}N_b + u(\omega)(B_{ba}N_b - B_{ab}N_a)
\end{array} \right.
\end{equation}
Here, $N_a$ and $N_b$ are the numbers of atoms in states $a$ and $b$ (with $E_b > E_a$); $A_{ba}$ is the spontaneous emission rate from $b \to a$; $B_{ab}$ is the absorption rate from $a \to b$; $B_{ba}$ is the stimulated (or induced) emission rate from $b \to a$; $u(\omega)$ is the energy density of the radiation field at the frequency $\omega=(E_b-E_a)/\hbar$. The first part of the Einstein paper~\cite{Einstein 1917} deals with energy transformations and the $A$ and $B$ rates of absorption and emission for processes in an atom or molecule in equilibrium with the radiation in a cavity. Incidentally, in this paper, for the probability that an atom decay spontaneously from state $b \to a$, Einstein takes $dW=A_{ba}dt$, and he quotes radioactive $\gamma$ decay and Hertzian oscillators as physical analogues. In the second part of his work, he addresses the momentum conservation in the radiation process concluding that in the spontaneous emission process ({\it Ausstrahlung}) the atom should recoil with magnitude $h \nu/c$ in a direction "[...]determined only by `chance' " ~\cite{PO50} (Einstein introduced in this way an element of chance in Quantum Mechanics). Spontaneous emission may be understood as the result of action of the particle as a whole, an immanent cause, occurring even if the system is closed (notwithstanding the possible role of the zero-point field~\cite{Milonni}). By the contrary, stimulate emission ({\it Einstrahlung}) occurs when the atom is an open system, interacting with the medium~\cite{Cornille_2003}, the initial and final states in the transition are defined by an external variable (i.e., the incident electric field), and the distinction between closed systems and open systems explain to a certain extent the existence of two types of radiation. Stimulated emission can be reenforced (by means of "optical pumping") by making the input wave with intensity $I_{\nu}$ traverses an inverted medium ($N_2 > N_1$), so that the radiation decays (or amplifies) according to $I_{\nu}(z)=I_{\nu}(0) e^{-\alpha z}$, with $\alpha=(N_1-N_2)\lambda^2 g(\nu)/8 \pi n^2 t_{spont}$, with $\lambda$ the wavelength of the radiation, $t_{spont}$ the spontaneous lifetime for $2 \to 1$ transitions,$n$ is the medium index of refraction, and $g(\nu)$ the lineshape function. Stimulated emission does not conserve energy, since atoms are open systems in a radiant medium. When the atom is submitted to a beam of plane waves propagating within the divergence angle of the beam, the momentum of the atom can be changed by stimulated absorption by the atom of a photon from one plane wave and subsequent stimulated emission into another plane wave; despite the two photons involved in these two processes have the same energy, however they differ by their propagating direction, resulting in a gradient force that can pulls the atom into or out of the laser beam; there is violation of the action-to-reaction force. This effect is used in optical tweezers.

\section{CONCLUSION}

The purpose of this study is to examine how the action-reaction law is presented in literature, particularly in what concern mechanics, electrodynamics and statistical mechanics, and to offer a methodological approach in order to clarify the fundamental aspects of the problem, in particular suggesting that a third system  must be included in the analysis of forces, what we call here, for the sake of conciseness, the physical vacuum.
Furthermore, our procedure leads to a generalization of the general linear canonical momentum of a
body-particle in the framework of statistical mechanics. Theoretical
arguments and numerical computations suggest that Newton's third law
is not verified in out-of-equilibrium systems, due to an additional term,
an entropic gradient term, which must be in the particle's canonical momentum.
Although Noether's theorem guarantee the conservation of canonical
momentum, the action-equal-reaction principle can be restored in
nonequilibrium conditions only if a new force term, representing the
action of the medium on the particles, is taken into account.

\section*{Acknowledgments}

The author gratefully acknowledge partial financial support by the
Technical University of Lisbon and the Funda\c{c}\~{a}o para a Ci\^{e}ncia e Tecnologia, Portugal.

\bibliographystyle{apsrmp}


\begin{thebibliography}{99}
\expandafter\ifx\csname natexlab\endcsname\relax\def\natexlab#1{#1}\fi
\expandafter\ifx\csname bibnamefont\endcsname\relax
  \def\bibnamefont#1{#1}\fi
\expandafter\ifx\csname bibfnamefont\endcsname\relax
  \def\bibfnamefont#1{#1}\fi
\expandafter\ifx\csname citenamefont\endcsname\relax
  \def\citenamefont#1{#1}\fi
\expandafter\ifx\csname url\endcsname\relax
  \def\url#1{\texttt{#1}}\fi
\expandafter\ifx\csname urlprefix\endcsname\relax\def\urlprefix{URL }\fi
\providecommand{\bibinfo}[2]{#2}
\providecommand{\eprint}[2][]{\url{#2}}




\bibitem[{\citenamefont{Abraham}(1909)}]{Abraham1}
\bibinfo{author}{\bibnamefont{Abraham}, \bibfnamefont{Max}},
  \bibinfo{year}{1910}, \bibinfo{journal}{Rend. Circ. Matem. Palermo}
  \textbf{\bibinfo{volume}{28}}, \bibinfo{pages}{1}.

\bibitem[{\citenamefont{Abraham}(1910)}]{Abraham2}
\bibinfo{author}{\bibnamefont{Abraham}, \bibfnamefont{Max}},
  \bibinfo{year}{1910}, \bibinfo{journal}{Rend. Circ. Matem. Palermo}
  \textbf{\bibinfo{volume}{30}}, \bibinfo{pages}{33}.

\bibitem[{\citenamefont{Aharonov} \emph{et~al.}(1988)\citenamefont{Aharonov,
  Pearle, and Vaidman}}]{Aharonov_88}
\bibinfo{author}{\bibnamefont{Aharonov}, \bibfnamefont{Y.}},
  \bibinfo{author}{\bibfnamefont{P.}~\bibnamefont{Pearle}}, and
  \bibinfo{author}{\bibfnamefont{L.} \bibnamefont{Vaidman}},
  \bibinfo{year}{1988}, \bibinfo{journal}{Phys. Rev. A}
  \textbf{\bibinfo{volume}{37}}, \bibinfo{pages}{4052}.

\bibitem[{\citenamefont{Antoci and Mihich}(1998)}]{Antoci}
\bibinfo{author}{\bibnamefont{Antoci}, \bibfnamefont{S.}} and
  \bibinfo{author}{\bibfnamefont{L.}~\bibnamefont{Mihichi}},
  \bibinfo{year}{1998}, \bibinfo{journal}{Eur. Phys. J. D}
  \textbf{\bibinfo{volume}{3}}, \bibinfo{pages}{205}.

\bibitem[{\citenamefont{Ashkin} \emph{et~al.}(1986)\citenamefont{Ashkin,
  Dziedzic, Bjorkholm, and Chu}}]{Ashkin 1986}
\bibinfo{author}{\bibnamefont{Ashkin}, \bibfnamefont{A.}},
  \bibinfo{author}{\bibfnamefont{J.~M.} \bibnamefont{Dziedzic}},
   \bibinfo{author}{\bibfnamefont{J.~E.} \bibnamefont{Bjorkholm}}, and
  \bibinfo{author}{\bibfnamefont{S.} \bibnamefont{Chu}},
 \bibinfo{year}{1986}, \bibinfo{journal}{Opt. Lett.}
  \textbf{\bibinfo{volume}{11}}, \bibinfo{pages}{299}.

\bibitem[{\citenamefont{Baierlein}(1968)}]{Baierlein}
\bibinfo{author}{\bibnamefont{Baierlein}, \bibfnamefont{R.}},
  \bibinfo{year}{1968}, \bibinfo{journal}{Am. J. Phys.}
  \textbf{\bibinfo{volume}{36}}, \bibinfo{pages}{625}.

\bibitem[{\citenamefont{Barrett}(1995)}]{Barrett}
\bibinfo{author}{\bibnamefont{Barrett}, \bibfnamefont{Terence W.}},
  \bibinfo{year}{1995}, in \emph{\bibinfo{booktitle}{Electromagnestism: foundations, theory and applications}},
  edited by \bibinfo{editor}{\bibfnamefont{Terence W.}~\bibnamefont{Barrett}} and
  \bibinfo{editor}{\bibfnamefont{Dale M.}~\bibnamefont{Grimes}}
  (\bibinfo{publisher}{World Scientific, Singapore}), p. \bibinfo{pages}{419}.

\bibitem[{\citenamefont{Baum and Kritikos}(1995)}]{Baum}
\bibinfo{author}{\bibnamefont{Baum}, \bibfnamefont{Carl E.}} and
  \bibinfo{author}{\bibfnamefont{Haralambos N.}~\bibnamefont{Kritikos}},
  \bibinfo{year}{1995}, \emph{\bibinfo{title}{Electromagnetic symmetry}}
  (\bibinfo{publisher}{Taylor $\&$ Francis, Washington, DC}).


\bibitem[{\citenamefont{Becker}(1964)}]{Becker}
\bibinfo{author}{\bibnamefont{Becker}, \bibfnamefont{R.}},
  \bibinfo{year}{1964}, \emph{\bibinfo{title}{Electromagnetic Fields and Interactions}}
  (\bibinfo{publisher}{Dover Publications, New York}).

\bibitem[{\citenamefont{Bellan}(1986)}]{Bellan 1986}
\bibinfo{author}{\bibnamefont{Bellan}, \bibfnamefont{P.~M.}},
  \bibinfo{year}{1986}, \bibinfo{journal}{Phys. Rev. Lett.}
  \textbf{\bibinfo{volume}{19}}, \bibinfo{pages}{2383}.

\bibitem[{\citenamefont{Bessonov}(1999)}]{Bessonov_99}
\bibinfo{author}{\bibnamefont{Bessonov}, \bibfnamefont{E.~G.}},
  \bibinfo{year}{1999}, \eprint{physics/9902065}.

\bibitem[{\citenamefont{Bondi}(1986)}]{Bondi 1986}
\bibinfo{author}{\bibnamefont{Bondi}, \bibfnamefont{Hermann}},
  \bibinfo{year}{1986}, \bibinfo{journal}{Proc. Roy. Soc. Lond. A}
  \textbf{\bibinfo{volume}{405}}, \bibinfo{pages}{265}.

\bibitem[{\citenamefont{B\"{o}ttger}(2005)}]{Waser}
\bibinfo{author}{\bibnamefont{B\"{o}ttger}, \bibfnamefont{Ulrich}},
  \bibinfo{year}{2005}, in \emph{\bibinfo{booktitle}{Polar Oxides: Properties,
Characterizing, and Imaging}},
  edited by \bibinfo{editor}{\bibfnamefont{R.}~\bibnamefont{Waser}}, \bibinfo{editor}{\bibfnamefont{U.}~\bibnamefont{B\"{o}ttger}}, and
  \bibinfo{editor}{\bibfnamefont{S.}~\bibnamefont{Tiedke}}
  (\bibinfo{publisher}{Wiley-VCH Verlag GmbH $\&$ Co. KGaA, Weinheim}), p. \bibinfo{pages}{11}.

\bibitem[{\citenamefont{Boyer}(2005)}]{Boyer_05}
\bibinfo{author}{\bibnamefont{Boyer}, \bibfnamefont{Timothy.~H.}},
  \bibinfo{year}{2005}, \bibinfo{journal}{Am. J. Phys.}
  \textbf{\bibinfo{volume}{73}}, \bibinfo{pages}{1184}.

\bibitem[{\citenamefont{Boyer}(2006)}]{Boyer_06}
\bibinfo{author}{\bibnamefont{Boyer}, \bibfnamefont{Timothy.~H.}},
  \bibinfo{year}{2006}, \bibinfo{journal}{J. Phys.:Math. Gen.}
  \textbf{\bibinfo{volume}{39}}, \bibinfo{pages}{3455}.

\bibitem[{\citenamefont{Brandt}(1989)}]{Brandt 1989}
\bibinfo{author}{\bibnamefont{Brandt}, \bibfnamefont{E.~H.}},
  \bibinfo{year}{1989}, \bibinfo{journal}{Science}
  \textbf{\bibinfo{volume}{243}}, \bibinfo{pages}{349}.

\bibitem[{\citenamefont{Breitenberger}(1968)}]{Breitenberger}
\bibinfo{author}{\bibnamefont{Breitenberger}, \bibfnamefont{Ernst}},
  \bibinfo{year}{1968}, \bibinfo{journal}{Am. J. Phys.}
  \textbf{\bibinfo{volume}{30}}, \bibinfo{pages}{505}.

\bibitem[{\citenamefont{Brillouin}(1970)}]{Brillouin_70}
\bibinfo{author}{\bibnamefont{Brillouin}, \bibfnamefont{L.}},
  \bibinfo{year}{1970}, \emph{\bibinfo{title}{Relativity Reexamined}}
  (\bibinfo{publisher}{Academic Press, New York}).

\bibitem[{\citenamefont{Brito}(2004)}]{Brito}
\bibinfo{author}{\bibnamefont{Brito}, \bibfnamefont{Hector}},
  \bibinfo{year}{2004}, \bibinfo{journal}{Acta Astronautica}
  \textbf{\bibinfo{volume}{54}}, \bibinfo{pages}{547}.

\bibitem[{\citenamefont{Buenzli and Soto}(2008)}]{Soto_08}
\bibinfo{author}{\bibnamefont{Buenzli}, \bibfnamefont{Pascal R.}}, and
  \bibinfo{author}{\bibfnamefont{Rodrigo}~\bibnamefont{Soto}},
  \bibinfo{year}{2008}, \bibinfo{journal}{Phys. Rev. E}
  \textbf{\bibinfo{volume}{78}}, \bibinfo{pages}{020102}.

\bibitem[{\citenamefont{Buenzli}(2009)}]{Buenzli 2009}
\bibinfo{author}{\bibnamefont{Buenzli}, \bibfnamefont{Pascal R.}},
  \bibinfo{year}{2009}, \bibinfo{journal}{J. Phys.: Conf. Series}
  \textbf{\bibinfo{volume}{161}}, \bibinfo{pages}{012036}.

\bibitem[{\citenamefont{Calkin}(1971)}]{Calkin_71}
\bibinfo{author}{\bibnamefont{Calkin}, \bibfnamefont{M.~G.}},
  \bibinfo{year}{1971}, \bibinfo{journal}{Am. J. Phys.}
  \textbf{\bibinfo{volume}{39}}, \bibinfo{pages}{513}.

\bibitem[{\citenamefont{Carpenter}(1989)}]{Carpenter_89}
\bibinfo{author}{\bibnamefont{Carpenter}, \bibfnamefont{C.~J.}},
  \bibinfo{year}{1989}, \bibinfo{journal}{IEE Proc.}
  \textbf{\bibinfo{volume}{136}}, \bibinfo{pages}{101}.

\bibitem[{\citenamefont{Chaumet, Nieto-Vesperinas and Rahmani}(2009)}]{Richards}
\bibinfo{author}{\bibnamefont{Chaumet}, \bibfnamefont{P.~C.}},
  \bibinfo{author}{\bibfnamefont{M.}~\bibnamefont{Nieto-Vesperinas}}, and
  \bibinfo{author}{\bibfnamefont{A.}~\bibnamefont{Rahmani}},
  \bibinfo{year}{2009}, in \emph{\bibinfo{booktitle}{Nano-Optics and Near-Field Optical Microscopy}}, edited by
  \bibinfo{editor}{\bibfnamefont{Anatoly}~\bibnamefont{Zayats}} and
  \bibinfo{editor}{\bibfnamefont{David}~\bibnamefont{Richards}}
  (\bibinfo{publisher}{Artech House, Norwood, MA}), p. \bibinfo{pages}{22}.

\bibitem[{\citenamefont{Chow}(2006)}]{Tsai}
\bibinfo{author}{\bibnamefont{Chow}, \bibfnamefont{Tsai}},
  \bibinfo{year}{2006}, \emph{\bibinfo{title}{Introduction to Electromagnetic theory: A Modern Perspective}}
  (\bibinfo{publisher}{Jones and Bartlett Publishers, Sudburt MA}).

\bibitem[{\citenamefont{Clevelance}(1996)}]{Clevelance}
\bibinfo{author}{\bibnamefont{Clevelance}, \bibfnamefont{Blair~M.}},
  \bibinfo{year}{1996}, \bibinfo{journal}{Electric Spacecraft}
  \textbf{\bibinfo{volume}{24}}, \bibinfo{pages}{6}.

\bibitem[{\citenamefont{Cohen-Tannoudji}
  \emph{et~al.}(1987)\citenamefont{Cohen-Tannoudji, Dupont-Roc, and
  Grynberg}}]{Cohen}
\bibinfo{author}{\bibnamefont{Cohen-Tannoudji}, \bibfnamefont{Claude}},
  \bibinfo{author}{\bibfnamefont{Jacques} \bibnamefont{Dupont-Roc}}, and
  \bibinfo{author}{\bibfnamefont{Gilbert}~\bibnamefont{Grynberg}},
  \bibinfo{year}{1987}, \emph{\bibinfo{title}{Introduction \`{a} l'\'{e}lectrodynamique quantique}}
  (\bibinfo{publisher}{InterEditions$/$Editions du CNRS, Paris}), p. \bibinfo{pages}{64}.

\bibitem[{\citenamefont{Cornish}(1986)}]{Cornish_86}
\bibinfo{author}{\bibnamefont{Cornish}, \bibfnamefont{F.~H.~J.}},
  \bibinfo{year}{1986}, \bibinfo{journal}{Am. J. Phys.}
  \textbf{\bibinfo{volume}{54}}, \bibinfo{pages}{166}.

\bibitem[{\citenamefont{Cornille}(1999)}]{Cornille}
\bibinfo{author}{\bibnamefont{Cornille}, \bibfnamefont{Patrick}},
  \bibinfo{1999}, \bibinfo{journal}{Progress in Energy and
Combustion Science} \textbf{\bibinfo{volume}{25}}, \bibinfo{pages}{161}.

\bibitem[{\citenamefont{Cornille}(2003)}]{Cornille_2003}
\bibinfo{author}{\bibnamefont{Cornille}, \bibfnamefont{Patrick}},
  \bibinfo{year}{2003}, \emph{\bibinfo{title}{Advanced Electromagnetism and Quantum Vacuum}}
  (\bibinfo{publisher}{World Sicneitif, New Jersey}).

\bibitem[{\citenamefont{Corum, Dering, Desavento, and Donne}(1999)}]{Corum 1999}
\bibinfo{author}{\bibnamefont{Corum}, \bibfnamefont{James F.}},
  \bibinfo{author}{\bibfnamefont{John P.}~\bibnamefont{Dering}},
  \bibinfo{author}{\bibfnamefont{Philip}~\bibnamefont{Pesavento}}, and
  \bibinfo{author}{\bibfnamefont{Alexsana}~\bibnamefont{Donne}},
  \bibinfo{year}{1999}, \bibinfo{journal}{AIP Conf. Proc.}
  \textbf{\bibinfo{volume}{458}}, \bibinfo{pages}{1027}.

\bibitem[{\citenamefont{Costa de Beauregard}(1967)}]{Costa_67}
\bibinfo{author}{\bibnamefont{Costa de Beauregard}, \bibfnamefont{O.}},
  \bibinfo{year}{1967}, \bibinfo{journal}{Phys. Letters}
  \textbf{\bibinfo{volume}{24A}}, \bibinfo{pages}{177}.

\bibitem[{\citenamefont{Costa de Beauregard}(1987)}]{Graneau_87}
\bibinfo{author}{\bibnamefont{Costa de Beauregard}, \bibfnamefont{O.}},
  \bibinfo{year}{1993}, \bibinfo{journal}{Phys. Lett. A}
  \textbf{\bibinfo{volume}{183}}, \bibinfo{pages}{41}.

\bibitem[{\citenamefont{Christodoulides}(1988)}]{Christodoulides}
\bibinfo{author}{\bibnamefont{Christrodoulides}, \bibfnamefont{C.}},
  \bibinfo{year}{1988}, \bibinfo{journal}{Am. J. Phys.}
  \textbf{\bibinfo{volume}{56}}, \bibinfo{pages}{357}.

\bibitem[{\citenamefont{Darwin}(1920)}]{Darwin_20}
\bibinfo{author}{\bibnamefont{Darwin}, \bibfnamefont{C.~G.}},
  \bibinfo{year}{1920}, \bibinfo{journal}{Phil. Mag.}
  \textbf{\bibinfo{volume}{39}}, \bibinfo{pages}{537}.

\bibitem[{\citenamefont{Denisov}(2002)}]{Denisov_02}
\bibinfo{author}{\bibnamefont{Denisov}, \bibfnamefont{S.}},
  \bibinfo{year}{2002}, \bibinfo{journal}{Phys. Lett. A}
  \textbf{\bibinfo{volume}{296}}, \bibinfo{pages}{197}.

\bibitem[{\citenamefont{Di\'{o}si}(2002)}]{Diosi_02}
\bibinfo{author}{\bibnamefont{Di\'{o}si}, \bibfnamefont{Lajos}},
  \bibinfo{year}{2002}, \eprint{physics/0206038}.

\bibitem[{\citenamefont{Dirac}(1951)}]{Dirac}
\bibinfo{author}{\bibnamefont{Dirac}, \bibfnamefont{P.}},
  \bibinfo{year}{1951}, \bibinfo{journal}{Nature}
  \textbf{\bibinfo{volume}{168}}, \bibinfo{pages}{906}.

\bibitem[{\citenamefont{Dzubiella} \emph{et~al.}(2003)\citenamefont{Dzubiella,
  {L\"{o}wen}, and Likos}}]{Likos_03}
\bibinfo{author}{\bibnamefont{Dzubiella}, \bibfnamefont{J.}},
  \bibinfo{author}{\bibfnamefont{H.}~\bibnamefont{{L\"{o}wen}}}, and
  \bibinfo{author}{\bibfnamefont{C.~N.}~\bibnamefont{Likos}},
  \bibinfo{year}{2003}, \bibinfo{journal}{Phys. Rev. Lett.}
  \textbf{\bibinfo{volume}{91}}, \bibinfo{pages}{248301-1}.

\bibitem[{\citenamefont{Edmonds}(1978)}]{Edmonds}
\bibinfo{author}{\bibnamefont{Edmonds, Jr.}, \bibfnamefont{James D.}},
  \bibinfo{year}{1978}, \emph{\bibinfo{title}{Maxwell's eight equations as one quaternion}},
  \bibinfo{journal}{Am. J. Phys.} \textbf{\bibinfo{volume}{46}}, \bibinfo{pages}{430}.

\bibitem[{\citenamefont{Einstein and Laub}(1908)}]{Laub}
\bibinfo{author}{\bibnamefont{Einstein}, \bibfnamefont{A.}}, and
\bibinfo{author}{\bibfnamefont{Laub}~\bibnamefont{J.}},
  \bibinfo{year}{1908}, \bibinfo{journal}{Annls. Phys.}
  \textbf{\bibinfo{volume}{26}}, \bibinfo{pages}{541}.

\bibitem[{\citenamefont{Einstein}(1912)}]{Einstein 1912}
\bibinfo{author}{\bibnamefont{Einstein}, \bibfnamefont{A.}},
  \bibinfo{year}{1912}, \bibinfo{journal}{Vierteljahrsschrift f\"{u}r gerichtlich Medizin und \"{o}ffentliches Sanit\"{a}tswesen}
  \textbf{\bibinfo{volume}{44}}, \bibinfo{pages}{37}.

\bibitem[{\citenamefont{Einstein}(1917)}]{Einstein 1917}
\bibinfo{author}{\bibnamefont{Einstein}, \bibfnamefont{A.}},
  \bibinfo{year}{1917}, \bibinfo{journal}{Phys. Z.}
  \textbf{\bibinfo{volume}{18}}, \bibinfo{pages}{121}.

\bibitem[{\citenamefont{Einstein}(1920)}]{Einstein22}
\bibinfo{author}{\bibnamefont{Einstein}, \bibfnamefont{A.}},
  \bibinfo{year}{1920}, \emph{\bibinfo{title}{Aether und
Relativitaetstheorie}}
  (\bibinfo{publisher}{Springer, Berlin}).

\bibitem[{\citenamefont{El Naschie}(2007)}]{El Naschie_1}
\bibinfo{author}{\bibnamefont{El Naschie}, \bibfnamefont{M.~S.}},
  \bibinfo{year}{2007}, \bibinfo{journal}{Int. J. of Nonlinear Sc. and Numerical Simulation}
  \textbf{\bibinfo{volume}{8}}, \bibinfo{pages}{469}.

\bibitem[{\citenamefont{Feigel}(2004)}]{Feigel_04}
\bibinfo{author}{\bibnamefont{Feigel}, \bibfnamefont{A.}},
  \bibinfo{year}{2004}, \bibinfo{journal}{Phys. Rev. Lett.}
  \textbf{\bibinfo{volume}{92}}, \bibinfo{pages}{020404-1}.

\bibitem[{\citenamefont{Feynman}(1964)}]{Feynman}
\bibinfo{author}{\bibnamefont{Feynman}, \bibfnamefont{R.~P.}},
  \bibinfo{year}{1964}, \emph{\bibinfo{title}{The Feynman Lectures}}
  (\bibinfo{publisher}{Adison Wesley, Reading, MA}).

\bibitem[{\citenamefont{Fowles}(1980)}]{Fowles_80}
\bibinfo{author}{\bibnamefont{Fowles}, \bibfnamefont{Grant ~R.}},
  \bibinfo{year}{1980}, \bibinfo{journal}{Am. J. Phys.}
  \textbf{\bibinfo{volume}{48}}, \bibinfo{pages}{779}.

\bibitem[{\citenamefont{Giner} \emph{et~al.}(1995)\citenamefont{Giner,
  Sancho, and Mart\'{i}nez}}]{Giner 1995}
\bibinfo{author}{\bibnamefont{Giner}, \bibfnamefont{V.}},
  \bibinfo{author}{\bibfnamefont{M.} \bibnamefont{Sancho}}, and
  \bibinfo{author}{\bibfnamefont{G.} \bibnamefont{Mart\'{i}nez}},
 \bibinfo{year}{1995}, \bibinfo{journal}{Am. J. Phys.}
  \textbf{\bibinfo{volume}{63}}, \bibinfo{pages}{749}.

\bibitem[{\citenamefont{Ginzburg and Ugarov}(1976)}]{Ginzburg76}
\bibinfo{author}{\bibnamefont{Ginzburg}, \bibfnamefont{V.~L.}}, and
\bibinfo{author}{\bibfnamefont{Ugarov}~\bibnamefont{V.~L.}},
  \bibinfo{year}{1976}, \bibinfo{journal}{Sov. Phys. Usp.}
  \textbf{\bibinfo{volume}{19}}, \bibinfo{pages}{94}.

\bibitem[{\citenamefont{Ginzburg}(1989)}]{Ginzburg}
\bibinfo{author}{\bibnamefont{Ginzburg}, \bibfnamefont{V.~L.}},
  \bibinfo{year}{1989}, \emph{\bibinfo{title}{Applications of Electrodynamics in Theoretical Physics and Astrophysics}}
  (\bibinfo{publisher}{Gordon and Breach Science Publishers, New York}).

\bibitem[{\citenamefont{Ginzburg and Frolov}(2002)}]{Ginzburg87}
\bibinfo{author}{\bibnamefont{Ginzburg}, \bibfnamefont{V.~L.}} and
  \bibinfo{author}{\bibfnamefont{V.~P.}~\bibnamefont{Frolov}},
  \bibinfo{year}{2002}, \bibinfo{journal}{Sov. Phys. Usp.}
  \textbf{\bibinfo{volume}{30}}, \bibinfo{pages}{1073}.

\bibitem[{\citenamefont{Glen, Murad, and Davis}(2008)}]{GlenMurad}
\bibinfo{author}{\bibnamefont{Glen}, \bibfnamefont{Robertson ~A.}},
  \bibinfo{author}{\bibfnamefont{P.~A.}~\bibnamefont{Murad}}, and
  \bibinfo{author}{\bibfnamefont{Eric}~\bibnamefont{Davis}},
  \bibinfo{year}{2008}, \bibinfo{journal}{Energy Conversion and Management}
  \textbf{\bibinfo{volume}{49}}, \bibinfo{pages}{436}.

\bibitem[{\citenamefont{Goldhaber}(1989)}]{Goldhaber_89}
\bibinfo{author}{\bibnamefont{Goldhaber}, \bibfnamefont{Alfred~S.}},
  \bibinfo{year}{1989}, \bibinfo{journal}{Phys. Rev. Lett.}
  \textbf{\bibinfo{volume}{62}}, \bibinfo{pages}{482}.

\bibitem[{\citenamefont{Gordon}(1973)}]{Gordon}
\bibinfo{author}{\bibnamefont{Gordon}, \bibfnamefont{J.~P.}},
  \bibinfo{year}{1973}, \bibinfo{journal}{Phys. Rev.}
  \textbf{\bibinfo{volume}{A8}}, \bibinfo{pages}{14}.

\bibitem[{\citenamefont{Graham and Lahoz}(1975)}]{Lahoz}
\bibinfo{author}{\bibnamefont{Graham}, \bibfnamefont{G.~M.}} and
  \bibinfo{author}{\bibfnamefont{D.~G.}~\bibnamefont{Lahoz}},
  \bibinfo{year}{1979}, \bibinfo{journal}{Nature}
  \textbf{\bibinfo{volume}{253}}, \bibinfo{pages}{339}.

\bibitem[{\citenamefont{Graham and Lahoz}(1979)}]{Lahoz2}
\bibinfo{author}{\bibnamefont{Graham}, \bibfnamefont{G.~M.}} and
  \bibinfo{author}{\bibfnamefont{D.~G.}~\bibnamefont{Lahoz}},
  \bibinfo{year}{1979}, \bibinfo{journal}{Phys. Rev. Lett.}
  \textbf{\bibinfo{volume}{42}}, \bibinfo{pages}{1137}.

\bibitem[{\citenamefont{Graham and Lahoz}(1980)}]{Lahoz3}
\bibinfo{author}{\bibnamefont{Graham}, \bibfnamefont{G.~M.}} and
  \bibinfo{author}{\bibfnamefont{D.~G.}~\bibnamefont{Lahoz}},
  \bibinfo{year}{1980}, \bibinfo{journal}{Nature}
  \textbf{\bibinfo{volume}{285}}, \bibinfo{pages}{154}.

\bibitem[{\citenamefont{Graham and Lahoz}(1980)}]{Lahoz}
\bibinfo{author}{\bibnamefont{Graham}, \bibfnamefont{G.~M.}}, and
  \bibinfo{author}{\bibfnamefont{D.~G.}~\bibnamefont{Lahoz}},
  \bibinfo{year}{1980}, \bibinfo{journal}{Nature}
  \textbf{\bibinfo{volume}{285}}, \bibinfo{pages}{154}.

\bibitem[{\citenamefont{Gerjuoy}(1949)}]{Gerjuoy:49}
\bibinfo{author}{\bibnamefont{Gerjuoy}, \bibfnamefont{E.}},
  \bibinfo{1949}, \bibinfo{journal}{Am. J. Phys.} \textbf{\bibinfo{volume}{17}}, \bibinfo{pages}{477}.

\bibitem[{\citenamefont{Graneau}(1982)}]{Graneau_82}
\bibinfo{author}{\bibnamefont{Graneau}, \bibnamefont{Peter},
  \bibinfo{1982}, \bibinfo{journal}{Nature} \textbf{\bibinfo{volume}{295}}, \bibinfo{pages}{311}.

\bibitem[{\citenamefont{Graneau}(1987)}]{Graneau_87}
\bibinfo{author}{\bibnamefont{Graneau}, \bibfnamefont{P.}},
  \bibinfo{year}{1987}, \bibinfo{journal}{J. Phys. D}
  \textbf{\bibinfo{volume}{20}}, \bibinfo{pages}{391}.

\bibitem[{\citenamefont{Graneau}(1994)}]{Graneau_94}
\bibinfo{author}{\bibnamefont{Graneau}, \bibfnamefont{Peter}},
  \bibinfo{year}{1994}, \emph{\bibinfo{title}{Ampere-Neumann Electrodynamics of Metals}}
  (\bibinfo{publisher}{Hadronic Press, Palm Harbour}).

\bibitem[{\citenamefont{Graneau and Graneau}(2001)}]{Graneau_01}
\bibinfo{author}{\bibnamefont{Graneau}, \bibfnamefont{Peter}} and
  \bibinfo{author}{\bibfnamefont{Neal}~\bibnamefont{Graneau}},
  \bibinfo{year}{2001}, \bibinfo{journal}{Phys. Rev. E}
  \textbf{\bibinfo{volume}{63}}, \bibinfo{pages}{058601}.

\bibitem[{\citenamefont{Greenberger} \emph{et~al.}(2007) \citenamefont{Greenberger, Erez,
 O. Scully, Svidzinsky and Zubairy}}]{PO50}
\bibinfo{author}{\bibnamefont{Greenberger}, \bibfnamefont{Daniel.~M.}},
  \bibinfo{author}{\bibfnamefont{Noam}~\bibnamefont{Erez}},
    \bibinfo{author}{\bibfnamefont{Marlan ~O.}~\bibnamefont{Scully}},
  \bibinfo{author}{\bibfnamefont{Anatoly ~A.}~\bibnamefont{Svidzinsky}},
  and \bibinfo{author}{\bibfnamefont{M.~Suhail}~\bibnamefont{Zubairy}},
  \bibinfo{year}{2007}, in \emph{\bibinfo{booktitle}{Progress in Optics}},
  \textbf{\bibinfo{volume}{50}}, edited by
  \bibinfo{editor}{\bibfnamefont{E.}~\bibnamefont{Wolf}}
  (\bibinfo{publisher}{Elsevier, New York}), pp. \bibinfo{pages}{275--327}.

\bibitem[{\citenamefont{Guala-Valverde and Achilles}(2008)}]{Valverde2}
\bibinfo{author}{\bibnamefont{Guala-Valverde}, \bibfnamefont{Jorge}}, and
\bibinfo{author}{\bibfnamefont{Achilles}~\bibnamefont{Ricardo}},
  \bibinfo{year}{2008}, \bibinfo{journal}{J. Grav. Phys.}
  \textbf{\bibinfo{volume}{2}}, \bibinfo{pages}{1}.

\bibitem[{\citenamefont{Guala-Valverde and Achilles}(2008)}]{Valverde1}
\bibinfo{author}{\bibnamefont{Guala-Valverde}, \bibfnamefont{Jorge}}, and
\bibinfo{author}{\bibfnamefont{Achilles}~\bibnamefont{Ricardo}},
  \bibinfo{year}{2008}, \bibinfo{journal}{Apeiron}
  \textbf{\bibinfo{volume}{15}}, \bibinfo{pages}{591}.

\bibitem[{\citenamefont{Haus and Penfield}(1968)}]{Haus_68}
\bibinfo{author}{\bibnamefont{Haus}, \bibfnamefont{H.~A.}} and
  \bibinfo{author}{\bibfnamefont{Penfield Jr.}~\bibnamefont{P.}},
  \bibinfo{year}{1968}, \bibinfo{journal}{Phys. Letters}
  \textbf{\bibinfo{volume}{26A}}, \bibinfo{pages}{412}.

\bibitem[{\citenamefont{Hnizdo}(1992)}]{Hnizdo_92}
\bibinfo{author}{\bibnamefont{Hnizdo}, \bibfnamefont{V.}},
  \bibinfo{year}{1992}, \bibinfo{journal}{Am. J. Phys.}
  \textbf{\bibinfo{volume}{60}}, \bibinfo{pages}{242}.

\bibitem[{\citenamefont{Ignatiev}(2007)}]{Ignatiev_07}
\bibinfo{author}{\bibnamefont{Ignatiev}, \bibfnamefont{A. Yu.}},
  \bibinfo{2007}, \bibinfo{journal}{Phys. Rev. Lett.} \textbf{\bibinfo{volume}{98}}, \bibinfo{pages}{101101}.

\bibitem[{\citenamefont{Ivanitski\u{i}} \emph{et~al.}(1991) \citenamefont{Ivanitski\u{i}, Medvinski\u{i},
and Tsyganov}}]{Tsyganov_91}
\bibinfo{author}{\bibnamefont{Ivanitski\u{i}}, \bibfnamefont{G.~R.}},
  \bibinfo{author}{\bibfnamefont{A.~B.}~\bibnamefont{Medvinski\u{i}}, and
  \bibinfo{author}{\bibfnamefont{M.~A.}~\bibnamefont{Tsyganov}},
  \bibinfo{year}{1991}, \bibinfo{journal}{Sov. Phys. Usp.}
  \textbf{\bibinfo{volume}{34}}, \bibinfo{pages}{289}.

\bibitem[{\citenamefont{Jammer}(1999)}]{Jammer}
\bibinfo{author}{\bibnamefont{Jammer}, \bibnamefont{Max}},
  \bibinfo{year}{1999}, \emph{\bibinfo{title}{Concepts of Force}}
  (\bibinfo{publisher}{Dover, New York}).

\bibitem[{\citenamefont{Jefimenko}(2000)}]{Jefimenko_1}
\bibinfo{author}{\bibnamefont{Jefimenko}, \bibfnamefont{Oleg ~D.}},
  \bibinfo{year}{2000}, \emph{\bibinfo{title}{Causality,
electromagnetic induction, and Gravitation: A different approach to
the theory of electromagnetic and gravitational fields}}
  (\bibinfo{publisher}{Electret Scientific Company, Star City}).

\bibitem[{\citenamefont{Jones and Richard}(1954)}]{Jones}
\bibinfo{author}{\bibnamefont{Jones}, \bibfnamefont{R.~V.}}, and
\bibinfo{author}{\bibfnamefont{J.~C.~S.}~\bibnamefont{Richard}},
  \bibinfo{year}{1954}, \bibinfo{journal}{Proc. R. Soc. A}
  \textbf{\bibinfo{volume}{455}}, \bibinfo{pages}{129}.

\bibitem[{\citenamefont{Keller}(1942)}]{Keller_42}
\bibinfo{author}{\bibnamefont{Keller}, \bibnamefont{J. M.}},
  \bibinfo{1942}, \bibinfo{journal}{Am. J. Phys.} \textbf{\bibinfo{volume}{10}}, \bibinfo{pages}{302}.

\bibitem[{\citenamefont{Khvorostenko}(1992)}]{Khvorostenko}
\bibinfo{author}{\bibnamefont{Khvorostenko}, \bibfnamefont{N.P.}},
  \bibinfo{year}{1992}, \emph{\bibinfo{title}{Longitudinal electromagnetic waves}},
  \bibinfo{journal}{Russian Phys. J.} \textbf{\bibinfo{volume}{35}}, \bibinfo{pages}{223}.

\bibitem[{\citenamefont{Koz{\l}owski and Marciak-Koz{\l}owska}(2002)}]{Kozlowski}
\bibinfo{author}{\bibnamefont{Koz{\l}owski}, \bibfnamefont{M.}} and
  \bibinfo{author}{\bibfnamefont{J.}~\bibnamefont{Marciak-Koz{\l}owska}},
  \bibinfo{year}{2002}, \bibinfo{journal}{Lasers in Engineering}
  \textbf{\bibinfo{volume}{12}}, \bibinfo{pages}{281}.

\bibitem[{\citenamefont{Landau and Lifchitz}(1970)}]{Landau FT}
\bibinfo{author}{\bibnamefont{Landau}, \bibfnamefont{L.}} and
  \bibinfo{author}{\bibfnamefont{E.}~\bibnamefont{Lifchitz}},
  \bibinfo{year}{1970}, \emph{\bibinfo{title}{Th\'{e}orie des Champs}}
  (\bibinfo{publisher}{Editions MIR, Moscow}).

\bibitem[{\citenamefont{Landau and Lifshitz}(1987)}]{Landau2}
\bibinfo{author}{\bibnamefont{Landau}, \bibfnamefont{L.~D.}} and
  \bibinfo{author}{\bibfnamefont{Lifshitz}~\bibnamefont{E.~M.}},
  \bibinfo{year}{1987}, \emph{\bibinfo{title}{Fluid Mechanics}}
  (\bibinfo{publisher}{Pergamon, Oxford}), \bibinfo{note}{2nd ed., Secs. 133 and 134}.

\bibitem[{\citenamefont{Landau and Lifshitz}(2007)}]{Landau TE}
\bibinfo{author}{\bibnamefont{Landau}, \bibfnamefont{L.~D.}} and
  \bibinfo{author}{\bibfnamefont{Lifshitz}~\bibnamefont{E.~M.}},
  \bibinfo{year}{2007}, \emph{\bibinfo{title}{Theory of Elasticity}}
  (\bibinfo{publisher}{Elsevier, Burlington, MA, 2007}).

\bibitem[{\citenamefont{Lavenda}(1974)}]{Lavenda}
\bibinfo{author}{\bibnamefont{Lavenda}, \bibfnamefont{B.~H.}},
  \bibinfo{year}{1974}, \bibinfo{journal}{Phys. Rev. A}
  \textbf{\bibinfo{volume}{9}}, \bibinfo{pages}{1}.

\bibitem[{\citenamefont{Lee}(1981)}]{Lee}
\bibinfo{author}{\bibnamefont{Lee}, \bibfnamefont{T.~D.}},
  \bibinfo{year}{1981}, \emph{\bibinfo{title}{Particle Physics and Introduction to Field theory}}
  (\bibinfo{publisher}{Harwood Academic Publishers, New York}).

\bibitem[{\citenamefont{Leff and Rex}(1990)}]{Leff}
\bibinfo{author}{\bibnamefont{Leff}, \bibfnamefont{Harvey ~S.}} and
  \bibinfo{author}{\bibfnamefont{Rex}~\bibnamefont{Andrew ~F.}},
  \bibinfo{year}{1990}, \emph{\bibinfo{title}{Maxwell's demon: entropy, information, computing}},
  edited by \bibinfo{editor}{\bibfnamefont{Harvey ~S.}~\bibnamefont{Leff}} and
  \bibinfo{editor}{\bibfnamefont{Andrew ~F.}~\bibnamefont{Rex}}
  (\bibinfo{publisher}{Adam Hilger, Bristol, MA}).

\bibitem[{\citenamefont{Mach}(1960)}]{Mach_01}
\bibinfo{author}{\bibnamefont{Mach}, \bibnamefont{Ernst}},
\emph{\bibinfo{title}{The Science of Mechanics}}, (\bibinfo{publisher}{Open Court Publishing Co., La Salle}\bibinfo{year}{1960}}),
\bibinfo{pages}{264ff}.

\bibitem[{\citenamefont{Maclay and Forward}(2004)}]{Forward2004}
\bibinfo{author}{\bibnamefont{Maclay}, \bibfnamefont{G.~Jordan}}, and
  \bibinfo{author}{\bibfnamefont{Robert L.}~\bibnamefont{Forward}},
  \bibinfo{year}{2004}, \bibinfo{journal}{Foundations of Physics}
  \textbf{\bibinfo{volume}{34}}, \bibinfo{pages}{477}.

\bibitem[{\citenamefont{Martins and Pinheiro}(2008)}]{Pinheiro 2008}
\bibinfo{author}{\bibnamefont{Martins}, \bibfnamefont{A.~A.}} and
  \bibinfo{author}{\bibfnamefont{Mario J.}~\bibnamefont{Pinheiro}},
  \bibinfo{year}{2008}, \bibinfo{journal}{Int. J. Theor. Phys.}
  \textbf{\bibinfo{volume}{47}}, \bibinfo{pages}{2706}.

\bibitem[{\citenamefont{Martins and Pinheiro}(2009)}]{Pinheiro 2009}
\bibinfo{author}{\bibnamefont{Martins}, \bibfnamefont{Alexandre ~M.}}, and
\bibinfo{author}{\bibfnamefont{Pinheiro}~\bibnamefont{Mario ~J.}},
  \bibinfo{year}{2009}, \bibinfo{journal}{Phys. Fluids}
  \textbf{\bibinfo{volume}{21}}, \bibinfo{pages}{097103}.

\bibitem[{\citenamefont{McDonald}(2006)}]{McDonald}
\bibinfo{author}{\bibnamefont{McDonald}, \bibfnamefont{Kirk T.}},
  \bibinfo{year}{2006}, \eprint{http://www.hep.princeton.edu/~mcdonald/examples/onoochin.pdf}.

\bibitem[{\citenamefont{Milgrom}(1983)}]{Milgrom_83}
\bibinfo{author}{\bibnamefont{Milgrom}, \bibfnamefont{M.}},
  \bibinfo{1983}, \bibinfo{journal}{Astrophys. J.} \textbf{\bibinfo{volume}{270}}, \bibinfo{pages}{365}.

\bibitem[{\citenamefont{Milonni}(1994)}]{Milonni}
\bibinfo{author}{\bibnamefont{Milonni}, \bibfnamefont{Peter~W.}},\bibinfo{pages}{23}
  \emph{\bibinfo{title}{The Quantum Vacuum--An Introduction to Quantum Electrodynamics}}, (\bibinfo{publisher}{Academic Press, Boston})\bibinfo{year}{1994}}.

\bibitem[{\citenamefont{Moffatt and Tokieda}(2008)}]{Moffatt 2008}
\bibinfo{author}{\bibnamefont{Moffatt}, \bibfnamefont{H.~K.}}, and
  \bibinfo{author}{\bibfnamefont{Tadashi}~\bibnamefont{Tokieda}},
  \bibinfo{year}{2008}, \bibinfo{journal}{Proc. Roy. Soc.}
  \textbf{\bibinfo{volume}{138A}}, \bibinfo{pages}{361}.

\bibitem[{\citenamefont{Newton}(2000)}]{Newton}
\bibinfo{author}{\bibnamefont{Hawking}, \bibfnamefont{Stephen}},\bibinfo{pages}{744}
  \emph{\bibinfo{title}{On the shoulders of giants: the great works of physics and astronomy}}, (\bibinfo{publisher}{Penguin Books, London})\bibinfo{year}{2002}.

\bibitem[{\citenamefont{Obara and Baba}(2000)}]{Baba_00}
\bibinfo{author}{\bibnamefont{Obara}, \bibfnamefont{Noriaki}} and
  \bibinfo{author}{\bibfnamefont{Mamoru}~\bibnamefont{Baba}},
  \bibinfo{year}{2001}, \bibinfo{journal}{Electronics and Communications in Japan, Part2}
  \textbf{\bibinfo{volume}{83}}, \bibinfo{pages}{31}.

\bibitem[{\citenamefont{Pearle}(1971)}]{Pearle_71}
\bibinfo{author}{\bibnamefont{Pearle}, \bibfnamefont{Philip}},
  \bibinfo{year}{1971}, \bibinfo{journal}{Phys. Rev. D}
  \textbf{\bibinfo{volume}{4}}, \bibinfo{pages}{1626}.

\bibitem[{\citenamefont{Pfeifer} \emph{et~al.}(2007)\citenamefont{Pfeifer,
  Nieminem, Heckenberg, and Rubinsztein-Dunlop}}]{Pfeifer 2007}
\bibinfo{author}{\bibnamefont{Pfeifer}, \bibfnamefont{Robert~N.~C.}},
  \bibinfo{author}{\bibfnamefont{Timo~A.}~\bibnamefont{Nieminem}},
  \bibinfo{author}{\bibfnamefont{Norman~R.}~\bibnamefont{Heckenberg}}, and
  \bibinfo{author}{\bibfnamefont{Halina}~\bibnamefont{Rubinsztein-Dunlop}},
  \bibinfo{year}{2007}, \bibinfo{journal}{Rev. Mod. Phys.}
  \textbf{\bibinfo{volume}{79}}, \bibinfo{pages}{1197}.

\bibitem[{\citenamefont{Pinheiro}(2002)}]{Pinheiro:02}
\bibinfo{author}{\bibnamefont{Pinheiro}, \bibfnamefont{Mario ~J.}},
  \bibinfo{year}{2002}, \bibinfo{journal}{Europhys. Lett.}
  \textbf{\bibinfo{volume}{57}}, \bibinfo{pages}{305}.

\bibitem[{\citenamefont{Pinheiro}(2004)}]{Pinheiro:04}
\bibinfo{author}{\bibnamefont{Pinheiro}, \bibfnamefont{M.~J.}},
  \bibinfo{year}{2004}, \bibinfo{journal}{Physica Scripta}
  \textbf{\bibinfo{volume}{70}}, \bibinfo{pages}{86}.

\bibitem[{\citenamefont{Pinheiro}(2007)}]{Pinheiro 2007}
\bibinfo{author}{\bibnamefont{Pinheiro}, \bibfnamefont{Mario ~J.}},
  \bibinfo{year}{2007}, \bibinfo{journal}{Physics Essays}
  \textbf{\bibinfo{volume}{20}}, \bibinfo{pages}{267}.

\bibitem[{\citenamefont{Plonsey and Collin}(1961)}]{Plonsey}
\bibinfo{author}{\bibnamefont{Plonsey}, \bibfnamefont{R.}},
  \bibinfo{author}{\bibfnamefont{R.~E.} \bibnamefont{Collin}},
  \bibinfo{year}{1961}, \emph{\bibinfo{title}{Principles and Applications of electromagnetic Fields}}
  (\bibinfo{publisher}{McGraw-Hill, New York}).

\bibitem[{\citenamefont{Poincar\'{e}}(1900)}]{Poincare 00}
\bibinfo{author}{\bibnamefont{Poincar\'{e}}, \bibfnamefont{H.}},
  \bibinfo{year}{1900}, \bibinfo{journal}{Archives n\'{e}erlandaises
des sciences exactes et naturelles}
  \textbf{\bibinfo{volume}{5}}, \bibinfo{pages}{252}.

\bibitem[{\citenamefont{Poincar\'{e}}(1900)}]{Poincare_01}
\bibinfo{author}{\bibnamefont{Poincar\'{e}}, \bibfnamefont{Henry}},
  \bibinfo{year}{2003}, \bibinfo{journal}{Archives n\'{e}erlandaises des
sciences exactes et naturelles}
  \textbf{\bibinfo{volume}{5}}, \bibinfo{pages}{252}; and
 \emph{\bibinfo{title}{Science and Method}} (Dover, New York, 2003).

\bibitem[{\citenamefont{Podolsky}(1966)}]{Podolsky}
\bibinfo{author}{\bibnamefont{Podolsky}, \bibfnamefont{B.}},
  \bibinfo{year}{1966}, \bibinfo{journal}{Am. J. Phys.}
  \textbf{\bibinfo{volume}{34}}, \bibinfo{pages}{42}.

\bibitem[{\citenamefont{Popov}(2002)}]{Popov_02}
\bibinfo{author}{\bibnamefont{Popov}, \bibnamefont{V. L.}},
  \bibinfo{2002}, \bibinfo{journal}{Technical Physics} \textbf{\bibinfo{volume}{47}}, \bibinfo{pages}{1397}.

\bibitem[{\citenamefont{Provatidis}(2010)}]{Provatidis 2010}
\bibinfo{author}{\bibnamefont{Provatidis}, \bibfnamefont{Christopher G.}},
  \bibinfo{year}{2010}, \bibinfo{journal}{Engineering}
  \textbf{\bibinfo{volume}{2}}, \bibinfo{pages}{648}.

\bibitem[{\citenamefont{Ravaisson}(1999)}]{Ravaisson}
\bibinfo{author}{\bibnamefont{Ravaisson}, \bibnamefont{F\'{e}lix}},
\emph{\bibinfo{title}{De l'habitude m\'{e}taphysique et morale}}, (\bibinfo{publisher}{Quadrige/PUF, Paris}\bibinfo{year}{1999}),
\bibinfo{pages}{8}.

\bibitem[{\citenamefont{Rengui}(1996)}]{Rengui}
\bibinfo{author}{\bibnamefont{Rengui}, \bibfnamefont{Ye}},
  \bibinfo{year}{1996}, \bibinfo{journal}{Eur. J. Phys.}
  \textbf{\bibinfo{volume}{17}}, \bibinfo{pages}{265}.

\bibitem[{\citenamefont{Ritz}(1908)}]{Ritz_08}
\bibinfo{author}{\bibnamefont{Ritz}, \bibfnamefont{Walter}},
  \bibinfo{year}{1908}, \bibinfo{journal}{Annales de Chimie et de Physique}
  \textbf{\bibinfo{volume}{13}}, \bibinfo{pages}{145}.

\bibitem[{\citenamefont{Sakuray}(1967)}]{Sakuray}
\bibinfo{author}{\bibnamefont{Sakuray}, \bibfnamefont{J.~J.}},
  \bibinfo{year}{1987}, \emph{\bibinfo{title}{Advanced Quantum Mechanics}}
  (\bibinfo{publisher}{Addison-Wesley, Reading, MA}).

\bibitem[{\citenamefont{Saumont}(1968)}]{Saumont_68}
\bibinfo{author}{\bibnamefont{Saumont}, \bibfnamefont{R.}},
  \bibinfo{1968}, \bibinfo{journal}{Phys. Lett. A} \textbf{\bibinfo{volume}{28}}, \bibinfo{pages}{365}.

\bibitem[{\citenamefont{Scanio}(1975)}]{Scanio_75}
\bibinfo{author}{\bibnamefont{Scanio}, \bibfnamefont{Joseph- J.~G.}},
  \bibinfo{year}{1975}, \bibinfo{journal}{Am. J. Phys.}
  \textbf{\bibinfo{volume}{43}}, \bibinfo{pages}{258}.

\bibitem[{\citenamefont{Selac} \emph{et~al.}(1989)\citenamefont{Selak,
  Messerotti, and Zlobec}}]{Selak 1989}
\bibinfo{author}{\bibnamefont{Selak}, \bibfnamefont{S.}},
  \bibinfo{author}{\bibfnamefont{M.}~\bibnamefont{Messerotti}}, and
  \bibinfo{author}{\bibfnamefont{P.}~\bibnamefont{Zlobec}},
  \bibinfo{year}{1989}, \bibinfo{journal}{Astrophys. Space Sci.}
  \textbf{\bibinfo{volume}{158}}, \bibinfo{pages}{159}.

\bibitem[{\citenamefont{Shockley and James}(1967)}]{ShockleyJames}
\bibinfo{author}{\bibnamefont{Shockley}, \bibfnamefont{W.}},
\bibinfo{author}{\bibfnamefont{R.~P.}~\bibnamefont{James}},
  \bibinfo{year}{1967}, \bibinfo{journal}{Science}
  \textbf{\bibinfo{volume}{156}}, \bibinfo{pages}{542}.

\bibitem[{\citenamefont{Shockley}(1967)}]{Shockley_67}
\bibinfo{author}{\bibnamefont{Shockley}, \bibfnamefont{W.}} and
\bibinfo{author}{\bibfnamefont{James}~\bibnamefont{R.~P.}},
  \bibinfo{year}{1967}, \bibinfo{journal}{Phys. Rev. Lett.}
  \textbf{\bibinfo{volume}{18}}, \bibinfo{pages}{876}.

\bibitem[{\citenamefont{Shockley}(1968)}]{Shockley_68}
\bibinfo{author}{\bibnamefont{Shockley}, \bibfnamefont{W.}},
  \bibinfo{year}{1968}, \bibinfo{journal}{Phys. Rev. Lett.}
  \textbf{\bibinfo{volume}{20}}, \bibinfo{pages}{343}.

\bibitem[{\citenamefont{Skobel'tsyn}(1974)}]{Skobeltsyn}
\bibinfo{author}{\bibnamefont{Skobel'tsyn}, \bibfnamefont{D.~V.}},
  \bibinfo{year}{1974}, \bibinfo{journal}{Sov. Phys.-Usp.}
  \textbf{\bibinfo{volume}{16}}, \bibinfo{pages}{381}.

\bibitem[{\citenamefont{Slepian}(1949)}]{Slepian}
\bibinfo{author}{\bibnamefont{Slepian}, \bibfnamefont{J.}},
  \bibinfo{year}{1949}, \bibinfo{journal}{Electrical Engineering}
  \textbf{\bibinfo{volume}{68}}, \bibinfo{pages}{145};
\bibinfo{author}{\bibnamefont{Slepian}, \bibfnamefont{J.}},
  \bibinfo{year}{1949}, \bibinfo{journal}{Electrical Engineering}
  \textbf{\bibinfo{volume}{68}}, \bibinfo{pages}{245}.

\bibitem[{\citenamefont{Soker and Harpaz}(2004)}]{Harpaz 2004}
\bibinfo{author}{\bibnamefont{Soker}, \bibfnamefont{Noam}} and
  \bibinfo{author}{\bibfnamefont{Amos}~\bibnamefont{Harpaz}},
  \bibinfo{year}{2004}, \bibinfo{journal}{Gen. Rel. Grav.}
  \textbf{\bibinfo{volume}{36}}, \bibinfo{pages}{315}.

\bibitem[{\citenamefont{Tangherlini}(1975)}]{Tangherlini}
\bibinfo{author}{\bibnamefont{Tangherlini}, \bibfnamefont{Frank~R.}},
  \bibinfo{year}{1975}, \bibinfo{journal}{Phys. Lett. A}
  \textbf{\bibinfo{volume}{12}}, \bibinfo{pages}{139}.

\bibitem[{\citenamefont{Taylor}(1965)}]{Taylor}
\bibinfo{author}{\bibnamefont{Taylor}, \bibfnamefont{T.~T.}},
  \bibinfo{year}{1965}, \bibinfo{journal}{Phys. Rev.}
  \textbf{\bibinfo{volume}{137}}, \bibinfo{pages}{B467}.

\bibitem[{\citenamefont{Thirring}(1927)}]{Thirring}
\bibinfo{author}{\bibnamefont{Thirring}, \bibfnamefont{Walter}},
  \bibinfo{year}{1927}, \emph{\bibinfo{title}{Dynamical Systems and Field Theories}}
  (\bibinfo{publisher}{Springer, Oxford}), bibinfo{note}{3rd ed.}.

\bibitem[{\citenamefont{Trammel}(1964)}]{Trammel}
\bibinfo{author}{\bibnamefont{Trammel}, \bibfnamefont{G. T.}},
  \bibinfo{year}{1964}, \bibinfo{journal}{Phys. Rev.}
  \textbf{\bibinfo{volume}{134}}, \bibinfo{pages}{B1183}.

\bibitem[{\citenamefont{van Tiggelen and Rikken}(2004)}]{Tiggelen_04}
\bibinfo{author}{\bibnamefont{van Tiggellen}, \bibfnamefont{B.~A.}} and
  \bibinfo{author}{\bibfnamefont{G.~L.~J.~A.}~\bibnamefont{Rikken}},
  \bibinfo{year}{2004}, \bibinfo{journal}{Phys. Rev. Lett.}
  \textbf{\bibinfo{volume}{93}}, \bibinfo{pages}{268903}.

\bibitem[{\citenamefont{{Vuji\v{c}i\'{c}}}(2004)}]{Vujicic_1}
\bibinfo{author}{\bibnamefont{{Vuji\v{c}i\'{c}}}, \bibfnamefont{V.~A.}}, and
  \bibinfo{author}{\bibfnamefont{J.~H.}~\bibnamefont{He}},
  \bibinfo{year}{2004}, \bibinfo{journal}{Int. J. of Nonlinear Sc. and Numerical simulation}
  \textbf{\bibinfo{volume}{5}}, \bibinfo{pages}{283}.

\bibitem[{\citenamefont{Walker and Lahoz}(1975)}]{Lahoz1}
\bibinfo{author}{\bibnamefont{Walter}, \bibfnamefont{G.~B.}} and
  \bibinfo{author}{\bibfnamefont{D.~G.}~\bibnamefont{Lahoz}},
  \bibinfo{year}{1975}, \bibinfo{journal}{Nature}
  \textbf{\bibinfo{volume}{253}}, \bibinfo{pages}{339}.

\bibitem[{\citenamefont{Wang} \emph{et~al.}(1989)\citenamefont{Wang,
  Sevick, Mittag, Searles, and Evans}}]{Wang_02}
\bibinfo{author}{\bibnamefont{Wang}, \bibfnamefont{G.~M.}},
  \bibinfo{author}{\bibfnamefont{E.~M.}~\bibnamefont{Sevick}},
  \bibinfo{author}{\bibfnamefont{E.}~\bibnamefont{Mittag}},
  \bibinfo{author}{\bibfnamefont{Searles}~\bibnamefont{D.~J.}}, and
  \bibinfo{author}{\bibfnamefont{D.~J.}~\bibnamefont{Evans}},
  \bibinfo{year}{2002}, \bibinfo{journal}{Phys. Rev. Lett.}
  \textbf{\bibinfo{volume}{89}}, \bibinfo{pages}{050601}.

\bibitem[{\citenamefont{Wesley}(1996)}]{Wesley 1996}
\bibinfo{author}{\bibnamefont{Wesley}, \bibfnamefont{J.P.}},
  \bibinfo{year}{1996}, \emph{\bibinfo{title}{Classical Quantum Theory}}
  (\bibinfo{publisher}{Benjamin Wesley, Weiherdammstasse 24, 78176 Blumber, Germany, Blumberg}).

\end{thebibliography}

\newpage

\end{document}